# NON-REPUDIATION FOR VOIP COMMUNICATION IN UMTS AND LTE NETWORKS

MASTER THESIS

For the Degree of

MSc in Information Technology Engineering

UMUT CAN ÇABUK

(201210733)

Supervisors

Dr. Rune Hylsberg Jacobsen, Aarhus University Department of Engineering, Denmark

Dr. Frank Gerd Weber, Fraunhofer Institute for Secure Information Technology, Germany

You can contact the author at *umutcancabuk@{gmail,hotmail,yahoo}.com*

# Abstract

This thesis work presents an architectural design of a system to bring non-repudiation concept into the IP based digital voice conversations (VoIP) in LTE and UMTS networks, using electronic signatures, by considering a centralized approach. Moreover, functionalities and technical methods to support such a system are researched. Last but not least, ways to introduce this system as a public and commercial service are discussed.

Non-repudiation concept provided by electronic signatures and related cryptographic functions, as introduced in this study, allow using digital records of these voice conversations as legally binding statements or proofs likewise and even instead of traditional wet signatures.

The system is designed as a subsystem to IMS based 3G and 4G networks and maximum compatibility with current configurations, components and interfaces of these networks is intended. On the other hand non-repudiation is achieved by special signature, storage and verification units located in the IMS core network. Voice data is proposed to be processed in MRF unit of the IMS. Additionally, a USSD/USSI based special solution to initiate these signed calls is developed.

According to the proposed scheme; during a signed call, two unidirectional voice streams originating from two parties of the call, which are transferred in IP and UDP encapsulated RTP packages, are received by the signature unit and interweaved using their arrival times, so that they become a unified stream. Signature unit generates hashes of groups of received packages and signs them using PKI algorithms and applying hash/signature chaining to increase integrity protection and to empower non-repudiation. Then, it forwards packages and signature information to the storage unit. Storage unit keeps all the call records, signature data and metadata of these calls. Verification unit later gathers relevant data from the storage unit and performs voice record playing as well as signature validation, when requested by any of the parties and/or legal entities.

Other than its main intention, this signed call service is believed to be useful in many different industries in many different ways, hence flexibility and modularity are discussed carefully throughout the study. Possible extensions like complex user verification methods or ways to expand the system over conventional circuit switched networks like GSM and PSTN are also adverted.

Finally, an implementation that demonstrates the key role and functionality of the signature unit is made to examine practical possibility, usability and feasibility of the proposed architecture.

# Acknowledgements

*“The scientific man does not aim at an immediate result. He does not expect that his advanced ideas will be readily taken up. His work is like that of the planter for the future. His duty is to lay the foundation for those who are to come, and point the way. He lives and labors and hopes.”*

Nikola Tesla

(“Radio Power will Revolutionize the World" in Modern Mechanics and Inventions, July 1934)

I would like to thank my academic supervisor Dr. Rune Hylsberg Jacobsen from Aarhus University for his valuable support and continual motivation during my whole master’s degree education.

I would like to thank my practical supervisors Dr. Frank Gerd Weber and Ronald Marx from Fraunhofer SIT for their valuable technical guidance during this thesis project.

Finally, I would like to thank my father, mother and all my family and friends for their unlimited fondness and amity.

# Table of Contents

# 1. Introduction

In this first chapter, a detailed introduction to the study is made by stating the needs that lead to such a work, eventual expectations, pioneering ideas, and the contextual coverage.

Above all, this work aims to utilize the power of VoIP communication, LTE and UMTS networks and digital signatures to design a system to provide ease of life and IT security in businesses to people, companies and governments.

## 1.1. Motivation and Scope

Digital voice carried on packet based IP networks, Voice over IP (VoIP) is becoming the prevalent form of voice communication in commercial environments and its popularity will keep this increasing trend in the future. Another increasing trend for decades, is the use of mobile (cellular) telephony, especially 3G/4G networks, instead of the fixed line systems in any kind of voice communication. Furthermore, UMTS and LTE mobile networks are designated to be able to support VoIP communication during voice calls, so this provides another boost to use of digital voice in commercial and public communication [1]. Moreover, like VoIP, SIP and RTP already are the standard communication protocols in UMTS and LTE core networks. Mobile networks are also able to provide new functionalities to regular VoIP calls using their unique architecture and integrated application servers with the power of IMS in their backbone.

In the recent releases of UMTS and LTE many security issues like confidentiality or authorization are currently addressed, but non-repudiation and secure integrity protection are not well addressed and are still active research topics. During prior studies, Fraunhofer SIT has developed a VoIP Signatures Protocol that splits the digital voice stream in fixed-length intervals, chain these intervals and electronically sign them to ensure an attacker cannot cut out, exchange, reorder or manipulate parts of the call. Additionally this prevents any party of the call from denying that this call has happened nor manipulated. Implementing VoIP Signatures as a network service to UMTS and LTE networks would strengthen authenticity and integrity in VoIP communication and will foster legally binding agreements closed via mobile phones on the fly.

Purpose of this work is to adapt the VoIP Signatures Protocol or a similar mechanism for securing the voice calls, into UMTS and LTE networks using a centralized approach, which will allow providers to offer this system as a public or commercial service. Additionally, a centralized approach would minimize the user side requirements and maximize the usability of the system. Besides, this would make lawful considerations to be controlled easier. Study also aims to investigate possible threats and problems in this concepts and thereby shall come up with additional solutions or functionalities to support the main service. As a key point, an implementation is considered to be useful to demonstrate the functionality and feasibility of such a system. Finally, extension possibilities and neoteric scenarios for the further research and developments are included to the study to spread this motivation to the architects of next generation communication systems.

## 1.2. Outline of the Thesis

This thesis presents the details of a research study, which proposes a new system architecture (or a framework) and new functionalities, mostly using known and standardized components and technologies. However, there are also some concepts including core network technologies and use of electronic signatures, which are still under research and development. With reference to that, the thesis is structured in the following scheme:

This first chapter explains the starting point of the study and gives details about what is intended to be achieved as a result as well as the methodology used. The following chapter gives introductory information about network architecture of modern cellular networks, voice communication on the internet, use of electronic signatures and legal aspects of such a system, as a background.

The third chapter is where the reader will be informed about the developed theory. It explains the proposed architecture with details and sets light to the usage and operation of the system. Additionally, it introduces all the designed and selected unique technologies and components those build up the framework. Furthermore, ways to allow this system to be a commercial service are explained. The fourth chapter propose solutions to extend the introduced system architecture, so that it covers the communication networks which do not natively support this framework or which are not equipped with this system.

The fifth chapter introduces a simplified implementation of the project, which is actually a part of the functionality that would be installed in a real industrial implementation. Also in the same chapter, results of the implementation and comments about these results are given. These results are mostly interpretative outcomes about the proposed functionality and its usefulness in a real industrial deployment, rather than raw numeric data. Sixth chapter demonstrates ideas about possible further extensions of this work in the context of industrial needs. Additionally, it gives a lead to the future researchers by presenting a bunch of topics as an academic springboard. Last but not least, seventh chapter concludes the project by giving a summary about produced ideas, usage scenarios, their implementations and made efforts in general.

The rest of the chapters written, provide complementary information about the study, including references, figure index and abbreviations glossary. Finally, some useful information which may be hard to find for readers somewhere else are added as two appendices.

Please note that the thesis uses a technical language that contains many related abbreviations, so keep an eye on the glossary. More, please not be confused, in the text, the system under research is also called the framework, the architecture and the service. By the same token, this thesis work is also referred as the research, the study and the project.

## 1.3. Assumptions

The project introduced in this thesis, relies on a few important assumptions to develop its ideas.

The very first assumption is the service provider support. Any kind of industrial development based on this project would require support by mobile network operators. So, the main implementers would set a collaboration with these operators or operators would do the implementations themselves.

A fundamental assumption is the legal recognition of electronic/digital signatures. The system proposed, by its nature, requires equality between wet signatures and electronic signatures. It can be implemented only in the countries/regions which have that legal background.

In addition to the legal assumption stated above, again we have to assume, provider of such a system like a mobile operator should beforehand have the written consent to digitally sign the intended voice calls instead of the clients (callers and callees) using their credentials and/or other information.

Another assumption is the existence of voice communication through IP infrastructure in LTE and UMTS networks. Any cellular network, which do not provide packet based voice transport feature, even though they might have LTE or UMTS network, will not be able to run this system properly.

The last assumption is the mutual trust between parties. All users of this system; clients, companies, government institutions as well as the providers are assumed to trust each other, and explicitly declare that by accepting terms and conditions of such a system, beforehand. Because signatures are legally binding statements and the service introduced in this study must be regulated officially. Calls made by unknown or untrusted parties are, inherently, not subjects of such a system.

## 1.4. Goals

The main goals of the study can be gathered under two main title: academic and industrial.

From an academic point of view, this study proposes a consistent and comprehensive communication architecture with use of appropriate mechanisms including digital signatures and related interfaces to provide a centralized non-repudiation functionality for voice calls and also provides integration solutions to make the system compatible with other current systems and subsystems of mobile networks.

On the other hand, from an industrial point of view, the study aims to come up with a guidance document (as the thesis) for a system, which is feasible to deploy, easy to use, auditable, secure against frauds, compatible with current architecture and flexible for future developments.

Hence, contextual goals of the study that are derived from these two points can be listed as follows;

1. Bringing non-repudiation concept to the voice calls made over mobile (cellular) networks.
2. Digitally signing and storing the voice data from an on-going call in real-time and providing ability to track this recorded copy of the call afterwards.
3. Producing a centralized approach to minimize the client-side requirements.

4. Recommending appropriate human interfaces and interactions to make all functionalities easily usable by users.
5. Providing an architectural design, which is as compatible as possible with the current 3G/4G network components and interfaces.
6. Keeping the architecture as simple in order to make the system feasible and realistic for investors and to prevent conflicts with other current systems or services.
7. Building the system in a modular and flexible way to make it open for further improvements.
8. Considering legal requirements to handle the sensitive user data and to use signed voice recordings as legally binding statements.
9. Introducing methods for the designed system to allow providers to serve it as a value service.

## 1.5. Scenario

A detailed scenario is created before launching the project, in order to objectify the goals and detect the requirements to reach these goals, using a set of realistic use cases. The scenario consists of two operations: Signed call session and record browsing. Signed call session starts after a user command. The session includes the ongoing call on the foreground and signing plus recording processes on the background. Then the session ends again by a user prompt. Record browsing is a secondary passive service which is available to the parties of the system and provides limited access to the database of signed calls via an appropriate user interface. The purpose of this service is to let users use their former calls as legally binding proofs for official cases. Furthermore none of the parties of a signed call would be able to deny that this call is happened nor be able to claim it is manipulated without him knowing. An exemplary storyline is given as follows:

A customer or investor needs to call a bank or a stock exchange to give various accounting orders. In most situations, banks and governments need written statements and even signatures from both parties of such orders or agreements. Thus the customer must visit an office to sign related documents, which will cost him/her time and money. Using the system proposed in this study, he could do the same thing via a phone call. Since he could be anywhere, even on the go, when this need appears, it is assumed that he would prefer to use his mobile phone. This assumption would also increase usability and coverage of such a system.

The procedure of usage supposed to be like the following: A customer indicates his intention to sign the upcoming call he wanted to make by sending a command and number of the callee to the provider, which presumably is a mobile network operator (MNO) and thus the MNO starts the session after reception of the command (See chapter 3.3 for details). Here the customer is chosen to be the caller, however the system shall support operations vice versa. The MNO makes a call to the callee, if the callee answers he is informed about that call to be signed. If he accepts, then the MNO calls back to the caller and the signed call starts. Context of the voice conversation is transparent to the system, parties may talk on any topic, in any language and using any sound characteristic. During the call, conversation is treated as a bidirectional digital voice stream, so it is signed and recorded in real-time. Signing and recording processes end when one of the parties (caller or callee) hangs up. Right after hanging up, caller receives a message from MNO, which asks caller for his final consent to store this completed call. If caller changes his mind and states that, signed conversation is deleted. If he confirms, then signed conversation is stored in the database. Finally both parties receive a disclosure notification and a tracking number to get access and listen the call later when necessary, hence the signed call session terminates.

About second operation of the scenario; MNO is required to provide an interface (a website is a good example) for users to listen these signed calls afterwards. Presumably each registered user of this system is given a username and a password by MNO in return for a written registration statement. Registration process is a key point to make sure of users' identities, so only in-person applications with official documents should be accepted by MNOs. The storyline may continue as follows: Customer later sees that his order is not taken into consideration by the bank, or in another case he needs proofs of his investments to prove his tax obligations in the tax office. He uses the website prepared to reach signed calls database, makes login by giving his username and password. By this way he is able to see all signed calls he made (maybe only the ones in a predefined period). If there are specific ones he needs, he may use the related tracking numbers and can directly reach the intended records. He is able to reach and print information about the calls, however downloading the actual calls is application dependent and may be available only if legal regulations allow.

## 1.6. Previous Works

Most of the concepts used and introduced in this research, like awareness of non-repudiation in digital communications, commercial use of electronic signatures and pure IP-based telephony as a native voice call service on mobile networks are very new and still subject to further developments and refinements. Thus, even though these concepts are researched and partially standardized individually, very few study touches on combined use of these concepts.

This project is based on an active research topic in the Mobile Networks Department of Fraunhofer Institute for Secure Information Technology (SIT) Darmstadt/Germany, which can be considered as a sequel work of prior studies of the department. In this work, two former theses, which introduce correlative concepts about VoIP signatures are used as origin. Although this project proposes a unique centralized non-repudiation architecture, which also concerns about compatibility with current mobile networks and native support by these networks; signature and storage mechanisms introduced by Hett [2] and Lauer [3] are reused with required modifications.

Christian Hett, in his thesis, proposes a method to achieve non-repudiation for SIP based private (PBX) fixed line VoIP telephony networks by electronically signing the continuous voice data in real-time and in a peer-to-peer (P2P) approach. The method is based on chains of hashes, generated from the voice data, and chains of electronic signatures. Using same method, Lauer in his work, provided a realistic implementation of the introduced method for Android based smartphones as a modification of SIPDroid application. All other resources can be found in the References section. As an illuminative note, unlike Hett's and Lauer's studies, this work proposes a centralized approach to provide the non-repudiation feature, plus a native support for most VoIP enabled phones without using any application on the user side. Moreover, the system is designated mainly for public/commercial mobile (cellular) networks.

# 2. Theoretical Background

In order to design, develop or improve a system that will satisfy the needs appeared in the scenario given in Section 1.5, some major technologies and concepts must be known a priori. Most, if not all, important theoretical foundations of the proposed system and functionalities are briefly introduced in this chapter. These background information contain the concept of non-repudiation, legal requirements of such a system and infrastructures of VoIP communication in UMTS and LTE networks.

## 2.1. Non-Repudiation

Non-repudiation is a security concept in information and communication technologies, which simply involves authentication and integrity of messages sent from all attendants of a data transmission or conversation session [4]. Hence, none of the parties may claim sent or received data are modified without their consent, nor this conversations is never happened.

Use of trusted third parties, which make use of digital signatures is one of the most common approaches to provide non-repudiation. In this work, providers are considered to act as trusted third parties. This is because, they already have appropriate infrastructure to verify and authenticate their clients connected to the cellular network. Furthermore, all data traffic between clients is already transferred using providers' network medium. However some exceptions exist in case of roaming and interworking, these situations are analyzed in Chapter 4.

## 2.2. Legal Considerations

Any public or enterprise oriented industrial and/or commercial implementation based on this work, shall require various legal directives as mainstay to consider electronically signed phone calls as "legally binding" data. Since it is the place where this study is done, laws of the Federal Republic of Germany are taken into consideration. However, thanks to the international organizations like European Union, laws in many countries are similar in this field. In addition to the laws, special agreements between users and providers may be required to realize such a system. For examples see Chapter 4.

Use of electronic signatures is regulated by the Law Governing Framework Conditions for Electronic Signatures (Signatures Law – SigG, published 2001, amended 2005) in Germany [5]. According to the law, first of all, service providers, most likely MNOs, should obtain accreditation to create electronic signatures and certificates. Other than signing, the law also implies recording the signed data (in our case, voice data) for a pre-determined period, again providers are responsible for that. They are also responsible for keeping these signatures and recordings safe, as well as users' private information. Making a signed call should clearly be under users' initiative and parties of the call must be informed about the process. In addition, MNOs must provide opportunity to users to back out or confirm the signing process after it is initiated or ended. Moreover, by law, providers should enable access to the recorded data upon request of parties of the call and/or government institutions. More details on the regulations can be found on the law document mentioned.

Another fact is that, as laws in most countries imply, signatures given under duress are later accepted as invalid. Hence, when compared to handwritten signatures, electronic signatures to be used in phone calls are less prone to be given under duress. Because, the system can be built in a way that allows cancellation of the signing process since it will ask for a confirmation after.

Prior to start using the system, all users should make a written agreement with the provider that states user's explicit consent to the provider for creating signatures and signing the intended voice calls instead of him. He/she should also confirm his phone number (also called IMSI number, issued with his/her SIM Card) to be used in this operation. Additionally, depending on the implemented solution, user should give or be given a PIN code or a secret question or a more complicated credential like voice recognition sample in order to add user verification layers (See Chapter 3 for details) and also to reach the voice records database afterwards. The exact method of user verification and the delivery method for these credentials are implementation dependent and are up to the MNOs.

## 2.3. VoIP

Voice over IP (VoIP) is a general name for a methodology and a set of technologies to digitalize and transport the analog voice data using IP packages, on a packet switched (PS) network. Besides transition to the PS domain network, in order to apply advanced operations on the voice data, like digital filtering, compressing, hashing, signing etc. it must be digitalized and should be divided into workable segments. Theoretically, the context of VoIP does not imply or enforce any specific technology or standard other than IP.

VoIP, if not all in most applications, uses Real-Time Transport Protocol (RTP) packages encapsulated in UDP datagrams those are embedded inside IP packages for transferring the digital voice data. Occasionally TCP can also be used instead of UDP. Digital voice data can be coded using different methods called codecs. Figure 1 shows the VoIP packet content overview. Figure 2 shows RTP packet structure, what a RTP header contains and where the actual voice data is located. More information regarding RTP can be found on its standardization document RFC 3550, published by IETF [6].

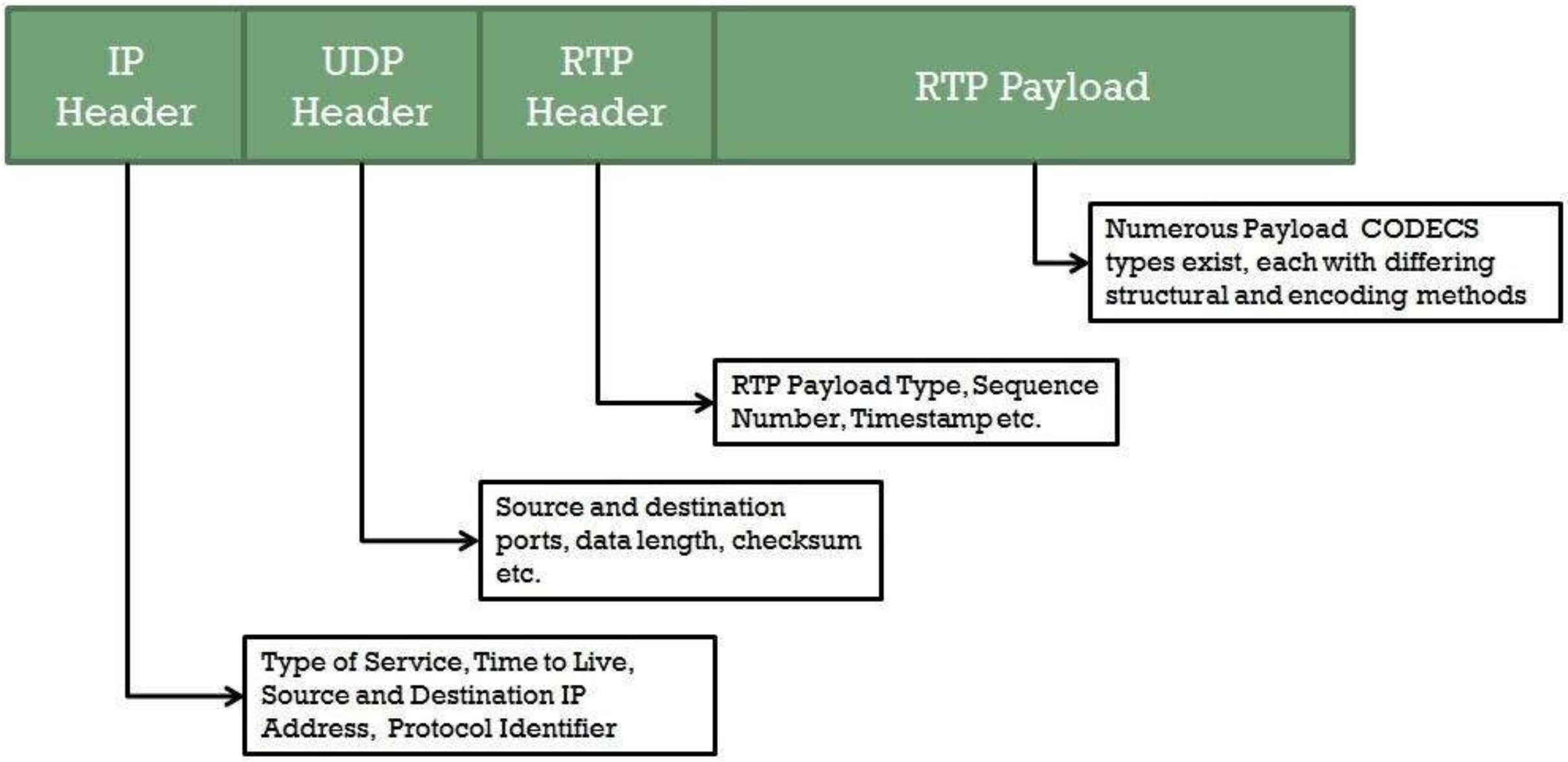

Figure 1: Voice packet structure in VoIP communication [f1].

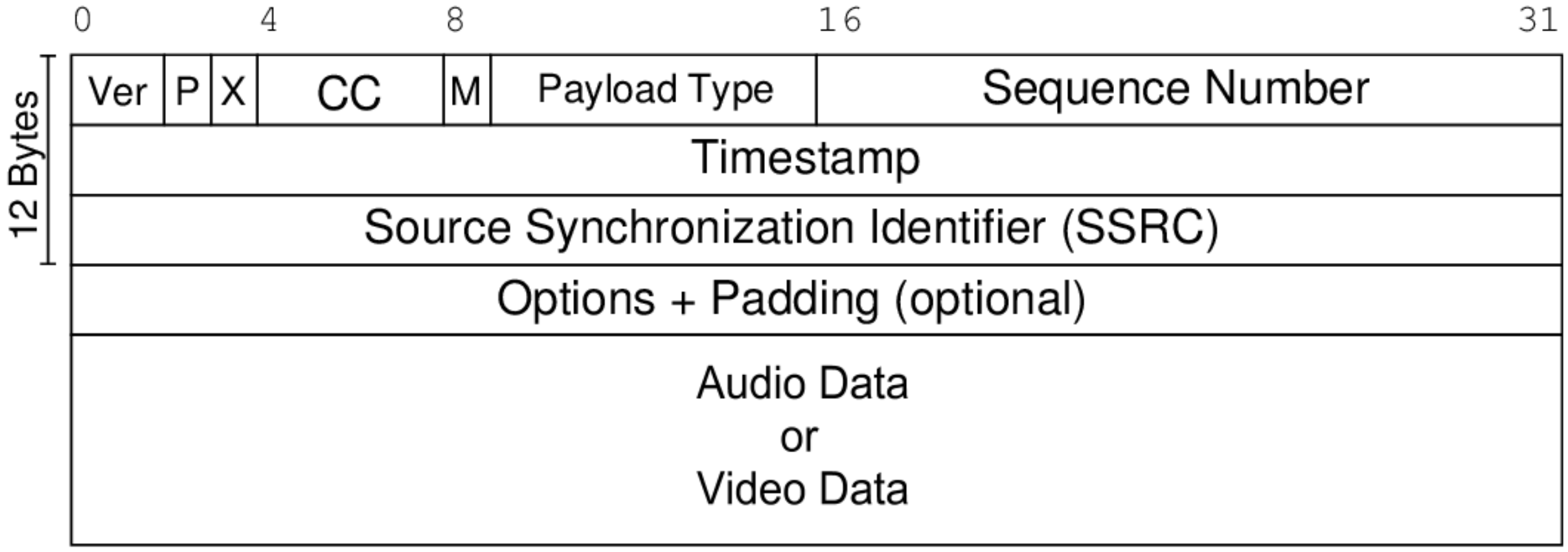

Figure 2: RTP packet internal structure [f2].

## 2.3.1. SIP

Session Initiation Protocol (SIP) is a signaling protocol, designed to control multimedia based communications made over IP networks. It is a text based application layer protocol, which can run on TCP, UDP, SCTP or TLS, and defined in the standardization document RFC 3261 by IETF [7].

SIP defines the messages that should be sent between endpoints of a communication session, necessary to establish, to terminate and to perform other operations of a call. SIP can be used for two-party (unicast) or multiparty (multicast) sessions. Other than VoIP, SIP features could be used to provide video conferencing, streaming multimedia distribution, instant messaging, presence information, file transfer, fax over IP and online games.

IP Based voice communication in 3G and 4G mobile networks powered by IMS (See Section 2.5 for more details), rely mostly on SIP in many of its interfaces, thus this study considers SIP as a fundamental protocol to take into account. Detailed information on request and response messages used in SIP can be found in the standardization document mentioned.

### 2.3.2. VoLTE

Voice over LTE (VoLTE) is an IMS based approach to bring VoIP calls to 4G/LTE networks, by carrying the voice traffic as data flows on the IP based LTE network directly using the data bearer without using circuit-switched fallback (CSFB). This approach eliminates the usage of CS voice domain and brings native support to voice conversations in PS domain. This feature is defined by GSMA in the document "PRD IR.92 - IMS Profile for Voice and SMS" [8].

Additionally, VoLTE supposed to provide high level of sound quality, which is important to use voice records as legally binding documents. 3GPP demands at least AMR Narrowband codec, but recommends AMR Wideband codec also referred as HD Voice. On the other hand, use of VoLTE with AMR can provide bandwidth efficiency by reducing the data overhead [9].

To use the opportunities of VoLTE, both user devices and operators should support that feature. Standardization of VoLTE is completed in 2013, so since the technology is very new, at the time of writing, the support in the market is very limited. First commercial VoLTE service launched in 2012, but the first full-featured (according to the standards) service launched in 2014 [10]. At the time of writing, the service exists only in a few countries. However, IP based voice is a rising trend in the industry and surely will be very popular soon. It can even be the dominant method for voice transfers of mobile networks in the near future. In this work, proposed functionalities and procedures designated to comply with current standardizations of VoLTE and/or other sub protocols implied by the IMS.

### 2.3.3. VoHSPA

Voice over HSPA (VoHSPA), like VoLTE, is an IMS based approach to bring VoIP calls to 3G/UMTS networks [11]. However, unlike LTE, in UMTS first intention was to make a circuit switched fallback when voice calls to be made. Then as newer approach, HSPA data network is used to digitalize and carry the voice data until to the existing CS infrastructure beyond the radio access network, this approach called CSoHSPA. After all, IMS structure is also implemented into UMTS architecture and thus VoHSPA provides the native VoIP support for UMTS.

Use of IMS model unified the VoIP support in LTE and UMTS. That also makes VoLTE and VoHSPA very similar in architecture, and makes them well-matched interoperable systems, even though they use different radio access networks.

## 2.4. LTE and UMTS Architectures

The system proposed in this study requires voice data to be carried in IP packages, and it is possible only in an IP based network architecture. Thus, legacy 2G GSM networks and earlier implementations of 3G UMTS networks prior to release 5, which handle voice calls only in circuit switched domain cannot be used to utilize this system. Furthermore, this data supposed to be processed in real-time during conversation in the network by an appropriate component, this requirement implies native multimedia handling in the core network. Hence, even though some concepts are introduced in earlier releases, any full or partial realization based on this study requires 3GPP Release 8 (or later) as the base version. See the next section for more details.

This work focuses more on the core networks rather than radio access networks. That's because, any implementation based on this study will mostly take place in the components of the core network of a MNO. Although, they have major differences in their radio access networks, namely UTRAN and e-UTRAN, as long as VoIP (VoLTE or VoHSPA) and IMS are implemented into the network, LTE and UMTS share a very similar core network structure. So, from the proposed system's point of view, RANs are transparent if they utilize PS domain for voice calls. Figure 3 illustrates the core network structure and interconnections of LTE network with support of UMTS and Figure 4 gives details about different RANs plus their connections to the core network as well as all the interconnections.

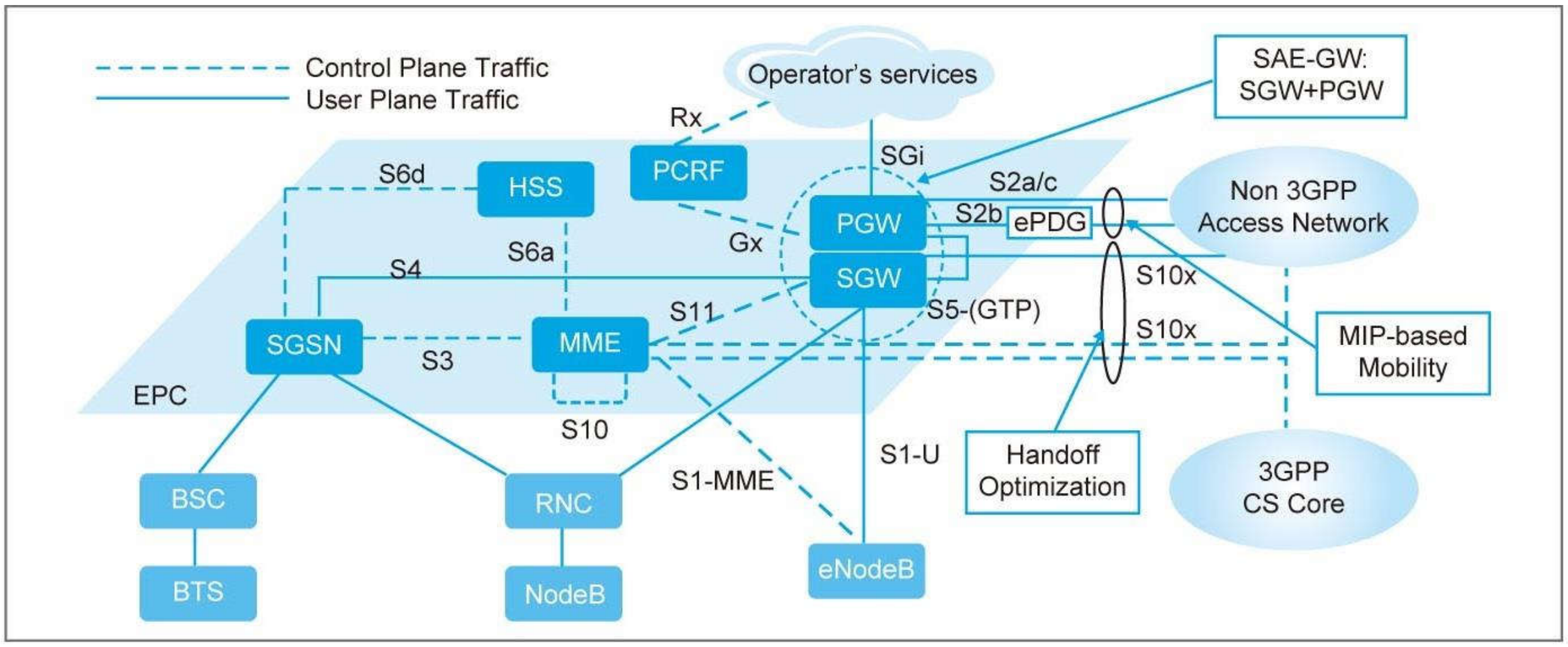

Figure 3: Route of user and control plane data in RANs and EPC with interfaces [f3].

In the Figure 3, while control plane traffic (dashed lines) stands for signaling messages, user plane traffic (straight lines) demonstrates the route of the actual voice data on the network. NodeB/eNodeB are base stations, PGW is the gateway for packet data, and the term "operator's IP services" mentions IMS and other interactional extensions like application servers.

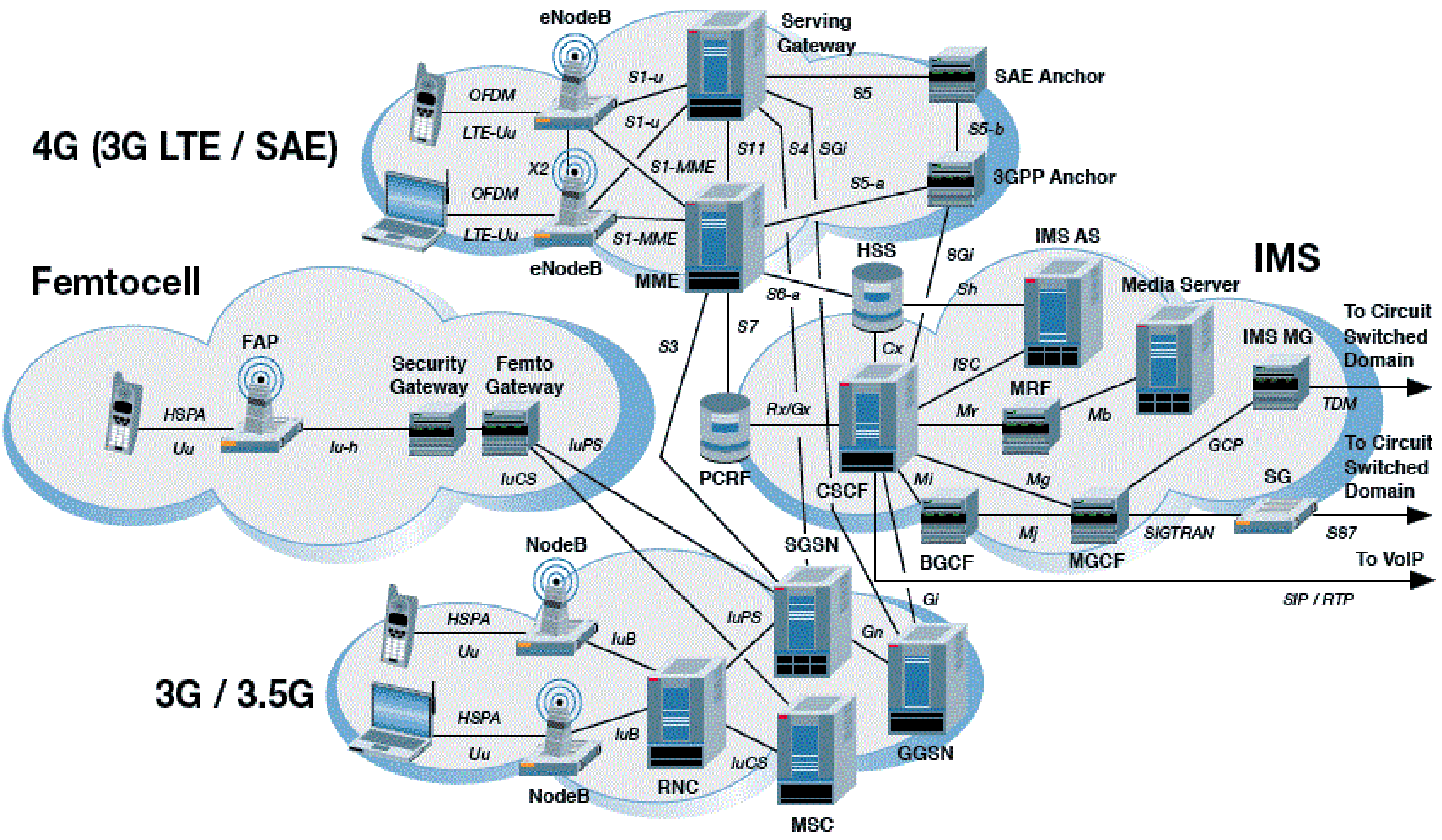

Figure 4: Interconnection between RANs and CN in LTE and UMTS [f4].

Figure 4 embodies how different RANs become transparent to the voice data in case of IMS is implemented in the MNO's core network. Here, Media Resource Function (MRF) of IMS is responsible for processing the voice data, when it is included to the network, it does not matter from which route the voice data is coming.

## 2.5. IMS

IP Multimedia - Core Network - Subsystem (IMS), is an architectural framework for providing multimedia services to contemporary telephony networks. Furthermore, another significant intention is to provide a standard for IP based voice conversations without making fallbacks on circuit switched domains. It is initially introduced in 3GPP Release 5 in a primitive form. In later releases, the framework is extended with new features like LTE support, emergency calls, SMS and USSD simulations etc. By the time of writing the most recent release is 3GPP Release 12 [12]. It is also considered as a recommended guideline in this work, even though Release 8 is taken as the basis. An architectural overview of IMS and its relation with other parts of the network is presented in Figure 5 below.

As some of the key components; Call Session Control Function (CSCF) is a role of a SIP server, responsible for determining a call's features and preferences. Policy and Charging Rules Functions (PCRF), formerly known as Policy Decision Function (PDF), is a server responsible for charging of a call depending on its various attributes. Media Resources Function (MRF) is a media server that is able to modify the voice data of a call, like applying echo cancellation or noise removal. More details about each component of IMS can be found elsewhere [13].

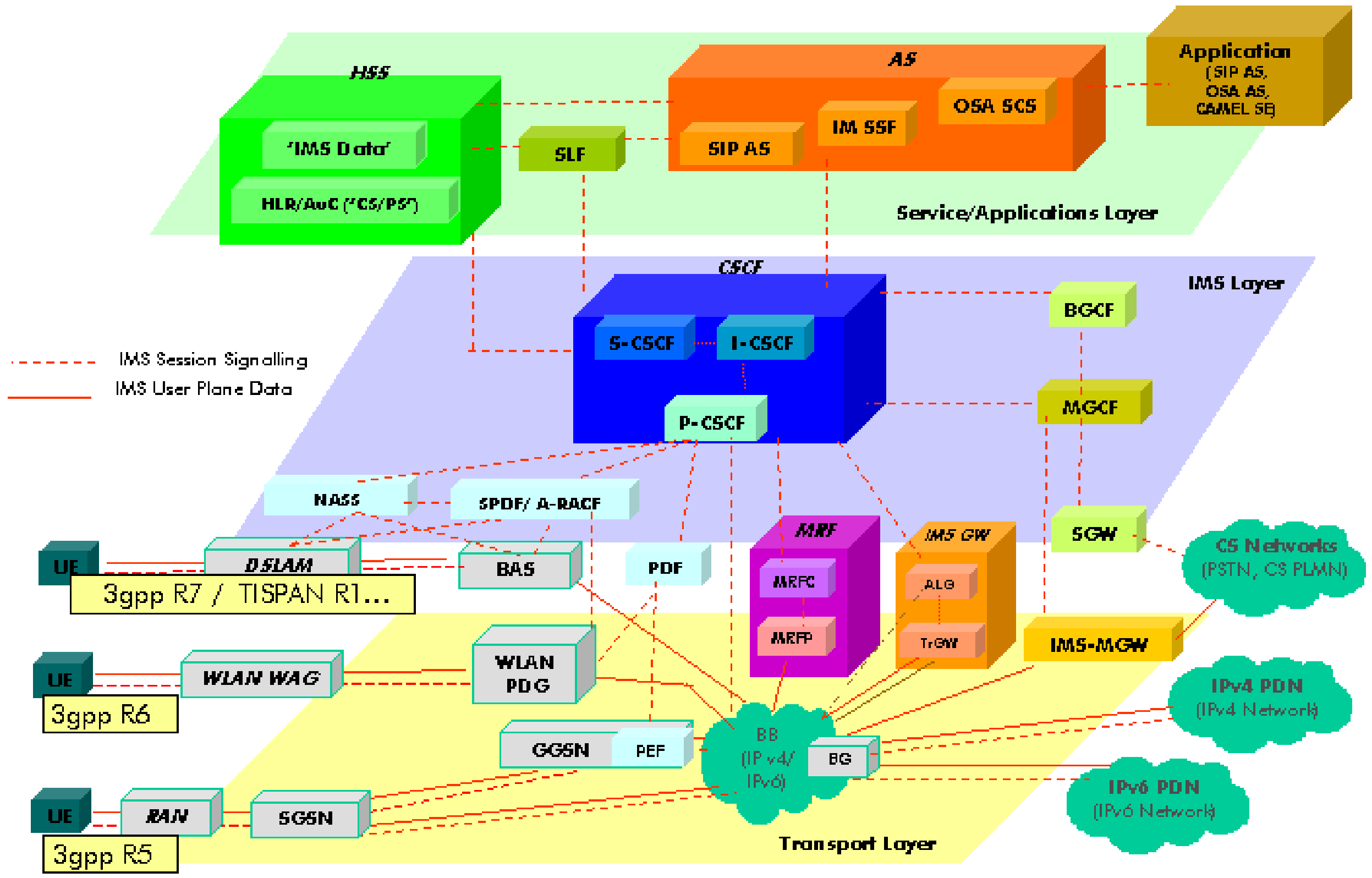

Figure 5: IMS based core network in LTE & UMTS, showing user and control plane data routes [f5].

Here in Figure 5, again straight lines represent the route of voice data. Assuming the VoLTE (or VoHPSA) case as the call handling scenario, the one and only component of IMS, which interacts with the voice data, is the Media Resources Function (MRF), in fact, Media Resources Function Processor (MRFP) of MRF. Hence, operations on the voice data proposed in this research (See Chapter 3 for details.), namely, signing the packages and marking them out for storing shall take place in this component. But, storing operation itself may be implemented in an additional component like a dedicated database server. Actually, there is no need to modify the raw voice data on the network, this is even undesirable, but making these operations on created copies of the incoming packages is intended instead. Therefore these operations can also be made in a separate application server on the packet data network of the MNO or even on the internet (See Section 6.3 for details). However, to provide singularity and consistency for voice data on the go, serial structure shouldn't be broken with such parallelism. This principle is implied by the concept of non-repudiation. In practice, this will prevent any intruder to manipulate farther copies of the floating voice data without knowledge of the provider MNO.

Figure 6 shows a detailed view of IMS Layer itself, including interfaces used in interconnections between its sub-components. Explanations of the interfaces are given in the Appendix A.

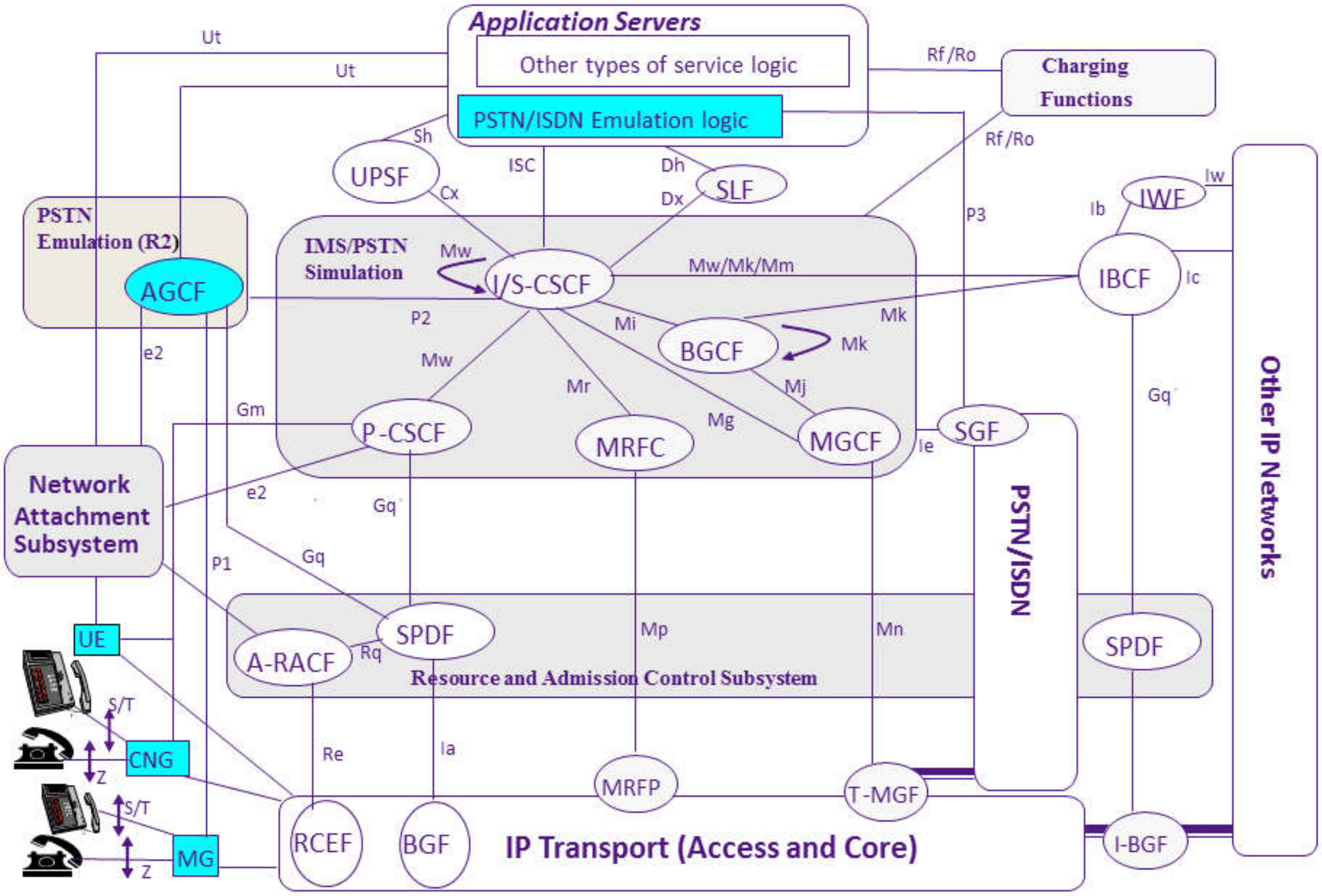

Figure 6: TISPAN and the IMS core with components and interfaces in LTE and UMTS networks [f6].

## 2.6. Digital Signatures and Certificates

Digital (or electronic) signatures are cryptographic techniques used to validate the authenticity and integrity of any kind of digital data, including e-mails, documents, messages, software, voice etc. In many countries, digital signatures have the same legal significance as the conventional wet signatures [14].

Public Key Infrastructure (PKI), which involves use of asymmetric cryptography powered by public and private key pairs, apart from encryption of the communication to hide the data from unwanted third parties, is also a common practice for digital signatures. Use of PKI in the context of non-repudiation is called Inverse Public Key Encryption (IPKE). RSA is a popular, and yet to be secure, public key cryptography standard [15]. It is also used and recommended within this study. Possible specialization opportunities to adopt the digital signatures to the proposed signed call framework are mentioned in the Section 3.2.

To electronically sign the digital documents using IPKE, first, a hash value is calculated from the subject data, then it is encrypted using sender's private key or related credentials. After, with a valid certificate, it is attached to the actual data and set or stored. Hence, the data is digitally signed in this way. To check the validity of the signature on a digital document to prove the genuineness of a document, first signature is decrypted using sender's public key. In parallel, hash is recalculated. If the decryption result from the signature matches the recalculated hash, this means the signature is genuine.

Digital certificates are electronic documents/files used to prove ownership of a public key. A digital certificate typically holds information about the key, its owner's identity and the digital signature of the certificate authority (CA) that has verified the certificate's contents are correct.

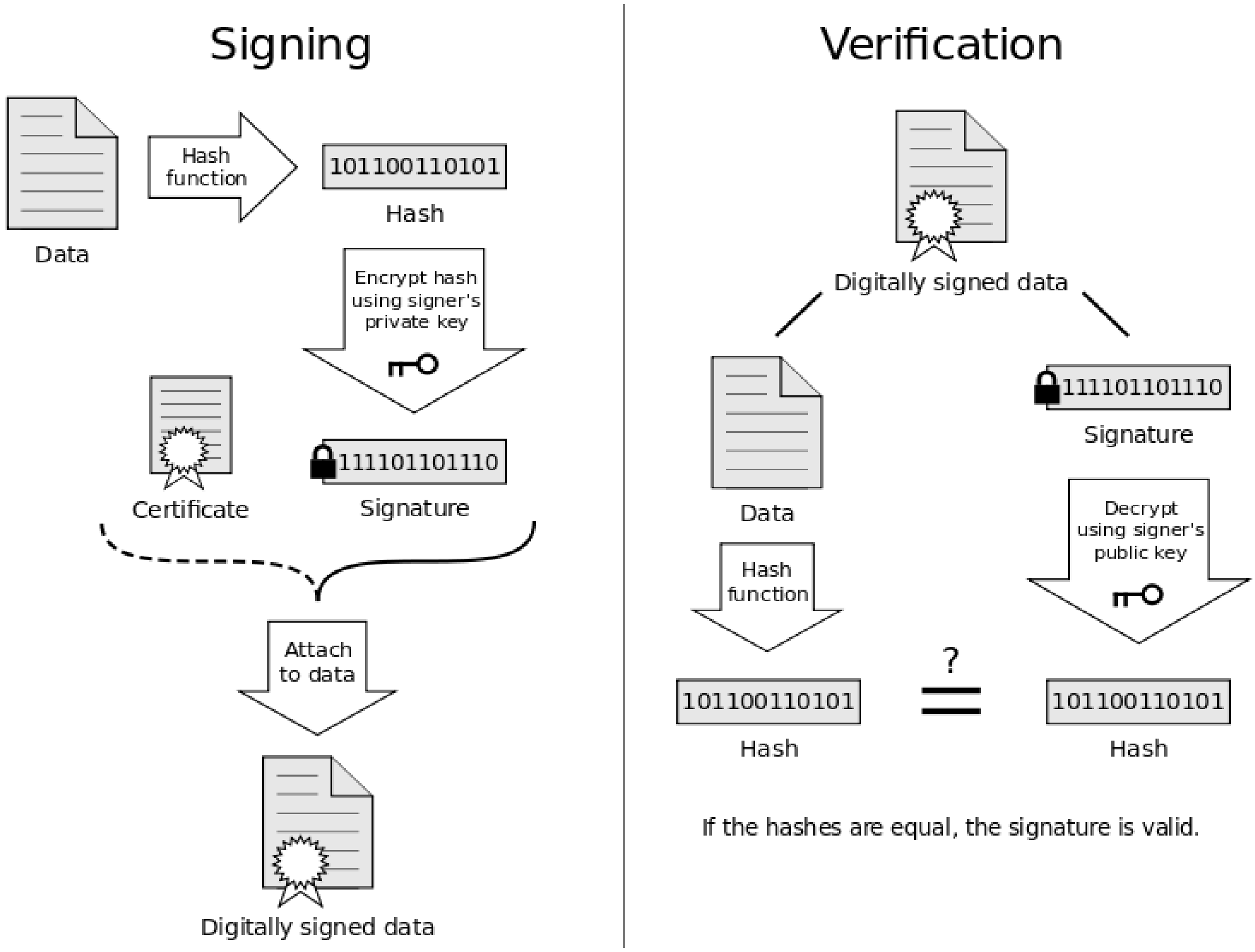

Figure 7: Creation and verification of digital signatures [f7].

## 2.7. USSD Service

Unstructured Supplementary Service Data (USSD) is a text based messaging protocol in GSM used to establish communication between mobile phones and the operators' special application servers over signaling channels. USSD messages support transfer of up to 182 alphanumeric characters. Unlike Short Message Service (SMS) messages, USSD messages create a real-time connection during a USSD session. The connection that allows a two-way exchange of a sequence of data, remains active until a termination message is sent or a timeout is reached. Because of that reason USSD is more responsive than services that use SMS [16].

USSD messages can be either UE or network initiated. Pull mode, handles mobile initiated USSD Requests and Push mode handles network initiated USSD requests or notifications. Push mode is usually used by the MNO to send, balance info, usage statistics, promotion and location notifications to the user. Pull mode is frequently used by the user to request informational services like news or weather, balance checking, subscription plan and tariff changing services. Within the scope of this work, also users initiate the service by sending a request. Incoming messages are displayed on the screen of the UE in a popup box immediately. However messages cannot be stored in the receiving phone. If the reply option is available, user can reply an incoming USSD push message.

Other than sequential messaging, USSD communication also allows the network to display a menu on the UE, which may have several different categories and options to be chosen by the user. These menu options may involve transactions regarding subscription services, phone settings, news, games etc. as well as special operator services like collect calls (reverse charging), scheduled calls and even the signed call service introduced in this thesis. Figure 8 illustrates an example USSD menu displayed on the phone screen with several options.

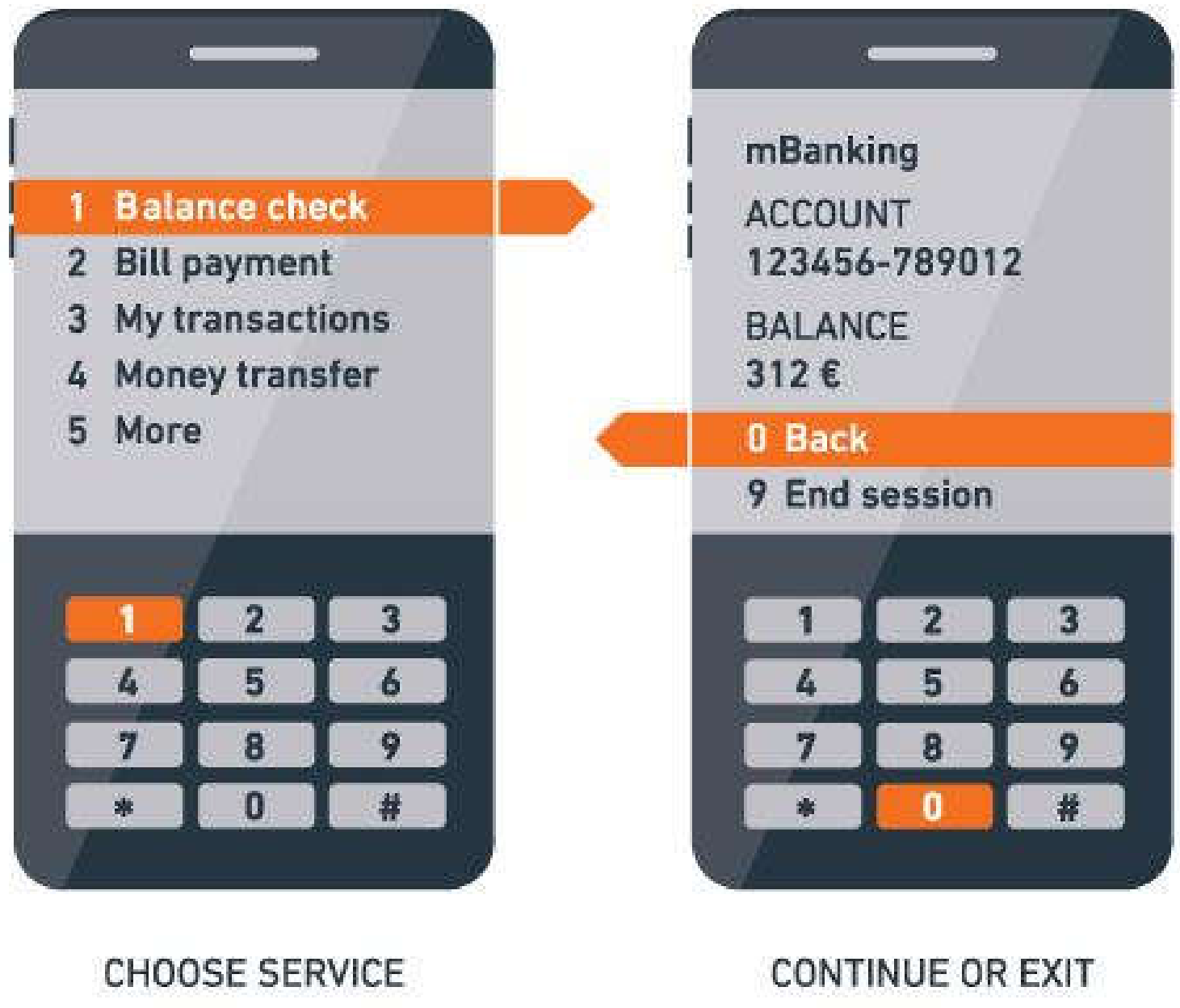

Figure 8: An example USSD menu on the phone screen [f8].

From the user side, messaging is initiated when user dials a code like *100# as a regular call. A typical USSD message starts with an asterisk/star sign (*) followed by digits that comprise commands, data or password. Groups of digits, for example commands and password, may be separated by additional asterisks between them. The message is terminated with a number/hash sign (#).

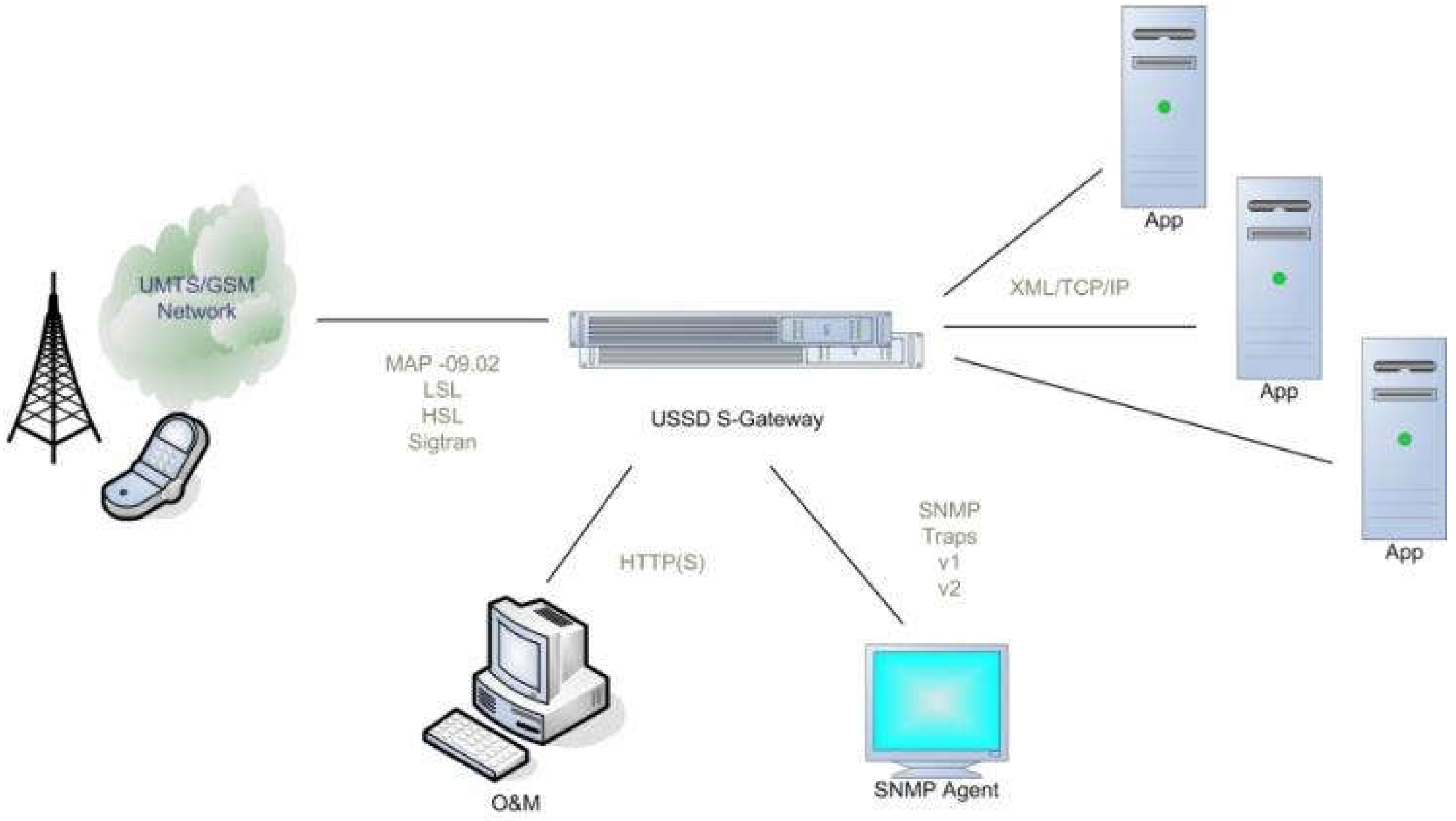

Figure 9: A typical USSD Network setup, gathered around USSD Gateway [f9].

A USSD message sent from the UE is handled by a USSD Gateway as can be seen in the Figure 9. Moreover, it is connected to several dedicated applications servers of the regarding MNO or even external service providers. There is no limitation on the applications running on these servers and they can be interconnected with other units of the core network like HLR/HSS, CSCF etc. However the data transfer between the UE and application servers are limited to alphanumeric texts only.

On the other hand, USSD is supported by almost any mobile phone just like SMS. While GSM and UMTS natively supports USSD, IP based flat LTE networks that utilize IMS, don't natively support it. To provide that support, USSD simulation interface (USSI) is recently standardized in the IMS framework. However implementation and use of USSI by MNOs is not a necessity, since most LTE phones are also capable of using CS domain with a basic CSFB, standard USSD service can be used without an issue. See Section 3.4 for more details.

# 3. A Non-Repudiative Voice Call Architecture for Mobile Networks

With regards to the initial goals of the study, an architecture (or scheme) is proposed as an outcome of the research, to satisfy the needs. During the development, maximum compatibility with the current modern cellular architecture and minimum effort to build is intended. So, the architecture introduced in this section is actually not a standalone solution, but a subsystem to the IMS equipped LTE and UMTS networks, explained in Sections 2.4 and 2.5.

In this thesis document, the word "system" refers to the proposed architecture including all related components, interfaces methods and functionalities. On the other hand, the word "service" is used to state all operations of the system as a commercial service.

## 3.1. Parties of the System

The system proposed, during its operation, is in relation with three different classes of real and legal entities: users, service providers and regulatory bodies. Not to be confused with parties of a call, who are only the users.

Regulatory bodies are usually the relevant government institutions, which make laws and ordinances to regulate the usage of such a system. Thus, they are responsible for assigning candidate service providers as trusted operators and allowing to them to run that service. Over and above, they are also the entities, whom these signed call records to be presented as legal evidences when necessary. This necessity may cover financial obligations, commercial disputes, statistical researches etc.

Service providers are the chosen trusted operators of such a system. Only providers selected by regulatory bodies can operate this system because of legal conditions. In this work, service providers are basically assumed to be mobile network operators (MNO), however in a future work or during the industrial implementation, system can be modified to be provided by other trusted third parties like private companies which can provide defined services in their own IMS-like public networks, as long as special agreements are made between MNOs and these trusted third parties. Providers responsible for signing call data using electronic signatures of users and storing them for a period which is defined in the laws regarding statute of limitations, in the country of deployment. Usually signature laws and frameworks, like EU Directive 1999/93/EC Framework for Electronic Signatures [17], do not mandate a specific time period for storage of recordings, but instead these periods are given in different laws depending on the field or sector. In many countries minimum of 10 years is a typical or recommended period [18].

Users are real and legal persons (entities), who want to have their phone conversations signed for future uses. In signed call both caller (who initiates the call) and callee (who receives the call) are users. In order to use the system, all users should register themselves to the provider. If any of the caller-callee pair is not registered, then a signed call shall not start and caller will be notified. Users exist in two categories: Real persons and legal persons. Here real persons are individuals who represent themselves and legal persons are companies or institutions who are represented by their official members or employees. Following cases explain some of the purposes in conversations between different user types:

- Real person to real person: Trade, debts, forensics…
- Real person to legal person: Shopping, account and stock orders, loans, service subscriptions…
- Legal person to legal person: Commerce, account and stock orders, legal reports…

In any case, any user can initiate the signed call process, there is no limitation about the initiator of a call. Furthermore, any person who initiated the call is able to verify the signature and listen the voice records regarding this call later. The example scenario introduced in Section 1.5 represents a signed call between a real person and a legal person, initiated by the real person.

## 3.2. Digitally Signing the Voice Data

Use of Public Key Infrastructure (PKI) based digital signatures are briefly explained in the Section 2.6, but within this work, since providers' core networks are presumably have a certain level of security since they are, supposed to be, unreachable to the intruders using public networks like internet. And since there is a centralized architecture which means all the interfaces, components and mediums are managed from the same authority, digitally signing the voice data is flexible in terms of format and methodology. This flexibility makes it possible to use Keyless Signature Infrastructure and signature methods like Merkle Signature Scheme, which are based on hash trees and one-time signatures [19]. Due to the closed structure of the cellular core networks, there is even no need to build a global infrastructure, but a localized single-use key generation infrastructure can be acceptable.

On the other hand, to keep the simplicity in the architecture and to be able to use the knowledge heritage from the previous works, as a design decision, in this work digitally signing the voice data refers to make use of hash generation and inverse public key cryptography (PKI) to provide non-repudiation and integrity to streaming voice data encapsulated in RTP packages. To achieve that, hash (or signature) chaining method is used prior to signing. This concept proposes inclusion of a previous signature as well as current hashes to be signed into a signature. Besides, there is no need to sign every single RTP packet alone. To reduce the delay and required processing power, signing process is proposed to be done for every interval, which is a set of consecutive RTP packages. Number of packages in an interval is not predefined, but because of size and duration values, 5 packet per interval is advised. Here, even if some packets are lost due to connectivity issues or unreliable UDP transmission, this won't lead to any problem. In such a case hash calculation is done only using received packages. Figure 10 shows how a group of packages are signed in a chained structure. Except for the first, every signature contains hash of the previous signature in addition to the packages. This is done to improve integrity protection on the stream [2] [3].

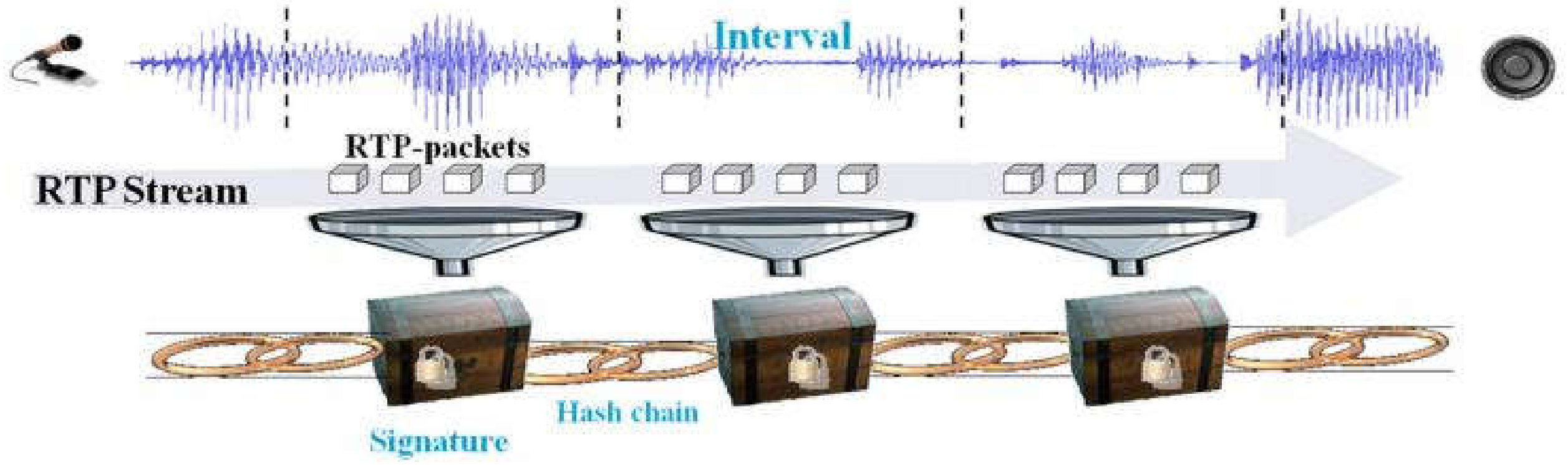

Figure 10: Signature scheme with hash chaining and packet losses [f10].

As a side note, here unlike former peer-to-peer solutions, in this centralized approach, MRFP interleaves incoming RTP packages from both directions prior to sign, and builds a plain full duplex stream like two people speaking in the same voice record. Signing process is done using the unified voice stream.

## 3.3. Multi-Factor User Authentication

Authentication and non-repudiation are closely related concepts of digital security. Since the system shall generate legally binding evidences from conversations, it is very important to be sure about genuineness of the users, who will sign the conversation. As partly mentioned in Chapter 2, LTE and UMTS networks already have standardized infrastructural components to authenticate their connected clients and to provide end-to-end security [20]. However MNOs can actually only check and authenticate the SIM/USIM Cards of users, and mobile phones in some cases. This is not enough in such a scenario introduced in Section 1.5, because verifying a USIM Card does not say anything about the person who uses that USIM Card. But, digital signatures should be unique for each user.

Multi-Factor User Authentication (MFA) is used to increase the certainty on the user identities by expanding the authentication coverage from device level to the personal level. It basically involves use of more than one authentication method consecutively. Methods can be summed up in three categories:

- Knowledge factors (things the user knows), i.e. PIN codes, passwords, secret questions etc.
- Possession factors (things the user has), i.e. chip cards, RFID appliances, USB dongles etc.
- Inherence factors (things the user is), i.e. finger prints, audio records, retina scans etc.

Even though knowledge factors might be weaker than possession and inherence factors in various situations; since one of the main academic challenges to be addressed in this work is to create a totally centralized approach, which eliminates the need of any kind of software or hardware installation on the user side; using knowledge factors is marked as the most useful method for the proposed system.

That's because most of the authentication methods, which use possession and inherence factors would require use of extra software like token interpreters or authenticating algorithms, and even extra hardware like card readers or USB connectors. These requirements are clearly in conflict with this work's attitude. To achieve our goal, simpler solutions derived from knowledge factors are chosen to be integrated to the system. Namely PIN codes, secret questions and one-time passwords. PIN codes are the most common solutions in similar situations used, however other options are also suitable in an industrial deployment.

## 3.4. Call Initiation Solutions

Since the call signing mechanism would not be a default functionality but an optional service, signed calls must be initiated differently than regular phone calls. As reference to the main goals of the project, users shall not be forced to use external applications. Hence, in order to initiate a signed call, three methods are offered. Any of these, and even all of these could be implemented in an industrial deployment. The choice is up to MNOs' preferences and to the local legal regulations.

### 3.4.1. Use of USSD/USSI Codes

Running USSD codes are done by typing a pre-defined command like *123# on the call screen and pressing YES/CALL/OK buttons on the mobile phone. This shall create session between the UE and the MNO. As reply to the code, MNO is able to send a text based menu to the UE, which appears automatically on the device screen like a notification. This menu shall request the number of the callee as a text input, after reception of the number, MNO can send another optional, but definitely recommended, question; the PIN code or the answer for a pre-determined secret question. These are specific for each user and required to prove the ownership of the used SIM/USIM/ISIM Card. So, with this prevention, any unauthorized access to the SIM Card is restricted and usage of the service is limited to the owner of the subscription, as laws imply.

After this phase, MNO automatically calls the callee. If he is available and if he answers, by a voice message on the line, he shall be informed about the incoming signed call and he shall be told that he accepts this if he stays on the line. Additionally a key press request can be implemented, for instance pressing 1 could carry on the call and pressing 2 could terminate the call. If that call is unsuccessful due to any reason, the caller is informed by a USSD notification and told to try later. If call is answered and accepted by the callee, MNO calls back the caller and connects the peers.

### 3.4.2. Use of USSD/USSI Direct Dialing

Another option is to type USSD codes in a longer form, which also covers the callee's number and the PIN (or answer). Such as *123*0123456789*0000# where 123 is the code for signing service, 0123456789 is the callee's number and 0000 is the PIN code of the caller. After sending this long command, the MNO shall not ask any more information in a reply also shall not display a menu, but shall start the session immediately. The rest of the procedure shall be the same as above. Both of this and the previous solutions can be implemented at the same time, so users can be free to choose the method they want. Similar methods are already in use to run services such as reverse charging calls, balance checking requests and operator specific setting downloads etc.

### 3.4.3. Use of SMS Messages

A different option to initiate a call is sending SMS messages instead of USSD codes. Since SMS messages, unlike USSD codes, do not create a real-time session for each use and are not transmitted momentarily, this option is not the primarily recommended way, but nice to implement as a replacement channel for the situations, when USSD communication is not available, such as temporary breakdowns and connectivity issues. In this case, caller sends callee's number and his PIN code in an SMS message to a pre-defined service number like 5555. The message shall contain a text like “0123456789 0000” without the quotes. Again 0123456789 is the callee's phone number and 0000 is the PIN code of the caller. The rest of the procedure shall be similar to the USSD based solutions. However, notifications from the MNO shall also be SMS messages, because when this mode is used, MNO shall assume USSD usage is not possible at that moment. This solution is presented as a reserve service and can be implemented in parallel with any or both of the USSD solutions. This work, especially in the implementation phase, mostly focuses on the USSD based solutions.

However, after a signed call is ended and the signature is confirmed by the caller, independent from the initiation method, an SMS notification which contains a success message and a tracking ID to reach the record later should be sent to the caller. This required because USSD notifications cannot be stored in the phone. Nevertheless, this is not the only way the user can reach his signed records. Any user shall have access to all his signed calls later via a web interface.

In Figure 11, a complete flow diagram from the system-functionality point of view, for a typical signed call is presented with all possible user or provider actions, except for faults presented in Section 3.9. In case of these faults, it can be assumed that the process will be canceled and all records will be erased if the call is started already. There, “Standby” is the situation for both parties are either idle or not using the system/service, however they can still be busy, i.e. having regular calls. The initiative for initiating a signed call belongs to the caller, but as can be seen, callee also has the opportunity to cancel the call prior to recording. If anyhow the call is cancelled (not same as terminated), process is cancelled and all voice records are deleted.

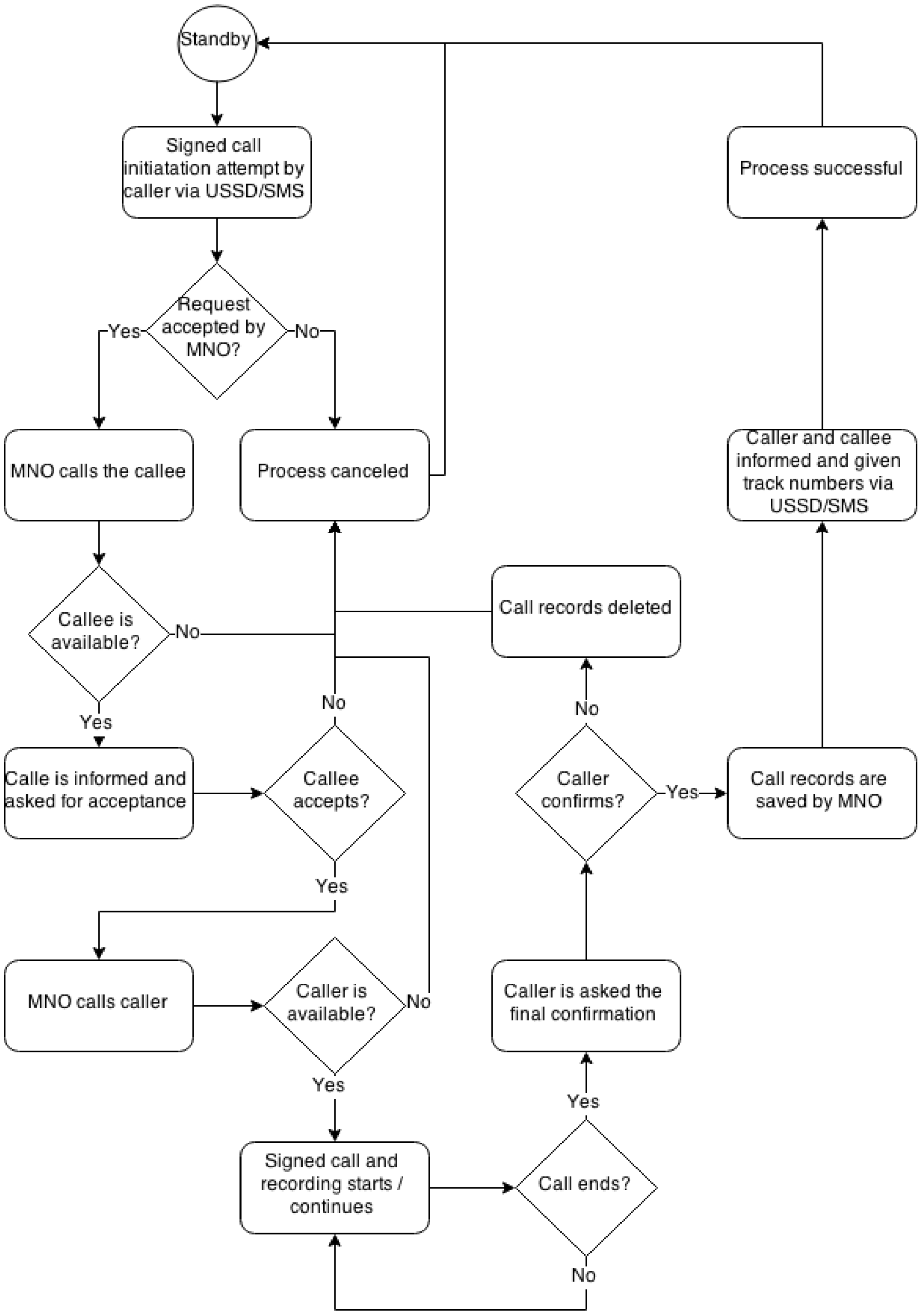

Figure 11: Complete flow diagram for a signed call process [f11].

## 3.5. Components and Interfaces

Components and interfaces that build up the system are given in this section. Since one of the goals of this study is maximum compatibility with the existing infrastructures, custom components and interfaces are used as few as possible. Most of the entities of LTE, UMTS and IMS in general, are reused. To avoid customization, new roles and functionalities are tried to be assigned in predefined or standardized units.

### 3.5.1. System Components

Even though there must be dedicated application servers to perform the intended functionalities, layers interconnections and interfaces of the network are preserved. Thus, the proposed system does not require any structural modification on the standardized IMS architecture, but some functional additions.

First of all, as long as VoIP calls (VoLTE or VoHSPA) and USSD communication are supported, there is no other special requirement on RANs. During a signed call, that part of the network is used as it is used during a regular VoIP call.

In the core network, there are several key components, which should be engineered particularly for the proposed system. Namely, a USSD application server to handle requests to initiate a signed call and to inform parties, a signing unit in the role of MRFP application server to perform the data processing operations, a storage server for voice dumps and signature data, a user credentials database to store information about registered users of the system and optionally an external application server for providing flexibility and extensibility in case of roaming or interworking scenarios introduced in Chapter 4.

Figure 12 shows the specially designed components of this signed call framework in relation with the Radio Access Network and the Core Network of a MNO's standard internal network. Red components build up the Centralized Non-Repudiation Core and perform the required major actions called Centralized Non-Repudiation Services. For more detail about these concepts, see Sections 3.8 and 3.9. Blue components are supplementary components which are fundamentally required to run and sustain the system in a secure way as intended. Additionally, the dashed lines show the data route for the signaling plane and the straight lines show the data route for the user plane.

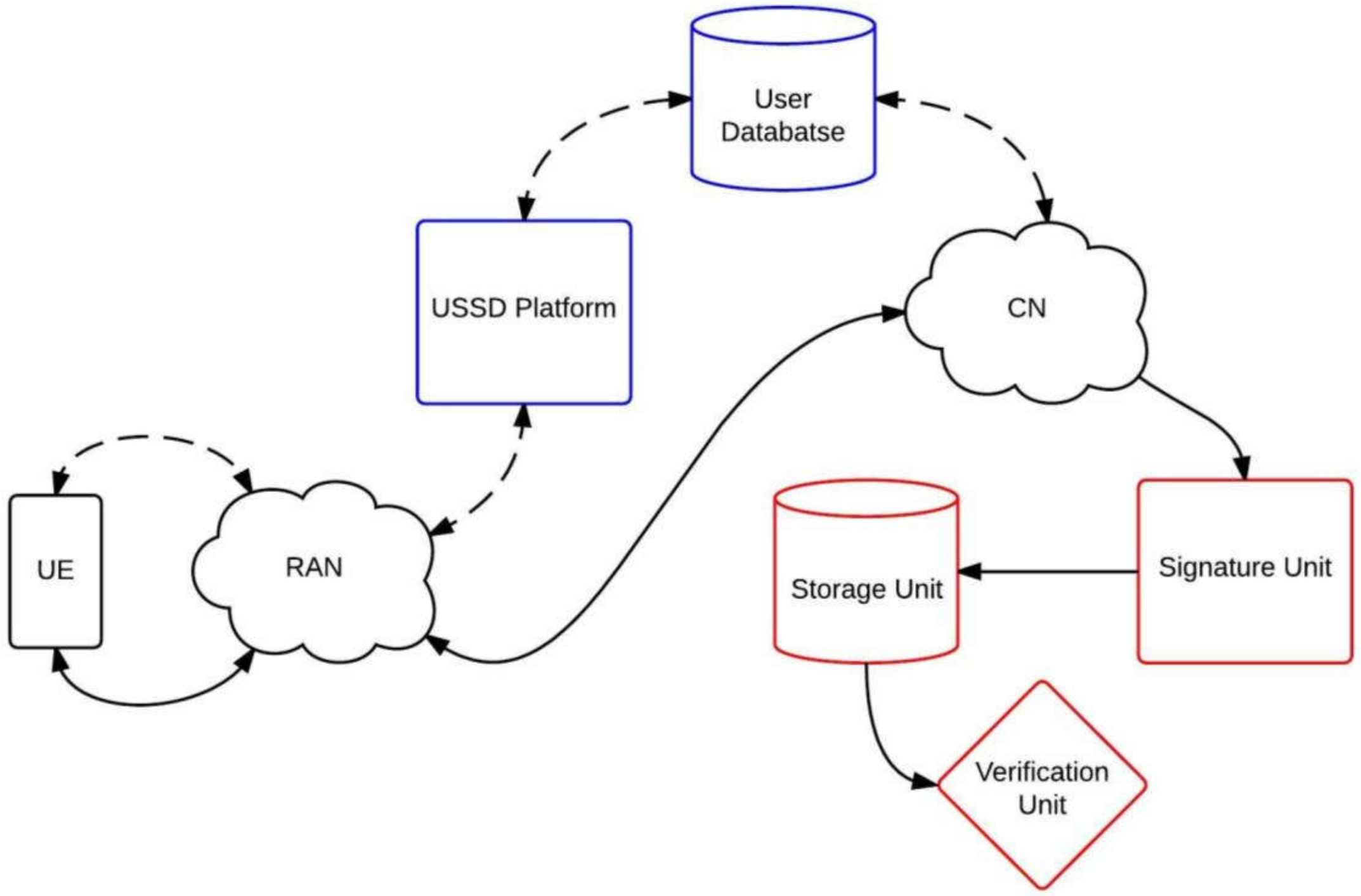

Figure 12: Designed components in relationship with the current networks [f12].

### 3.5.2. Cycle of Voice Data

In a mobile network, during a call, the journey of voice data, also called as user plane data, starts from one UE passes through RAN, CN, IP backbone (if available) and ends at the destination UE simultaneously in a bidirectional manner. When a signed call is set up, voice data is transmitted in the RAN like it is a regular VoIP call. On the other hand, when voice data reaches to the core network (EPC for the LTE case) then the procedure becomes special. PGW sends the incoming voice data to operator's IP backbone using SGi interface. Here MRF intercepts the packets with reception of a control plane invoke from CSCF to MRFC. Then MRFP, in real-time, makes copies of incoming packets as they generate the same voice stream like the actual conversation, then it allows each packet to continue its regular route to the next destination in the IP network. Again in real-time, MRFP signs the copies of the voice data and forwards them to the storage unit using a custom interface. To make it clear, actual voice data is not modified, but copies of them are. Although it can be done in another application server on a higher layer, using the MRF unit for the signing role is necessary to prevent any copies of the voice data floating in the network. This parallelism could allow manipulation of the voice data without parties of the conversation noticing it. More details about the signaling and user plane data communication are presented as step-by-step in the Apendix B.1.

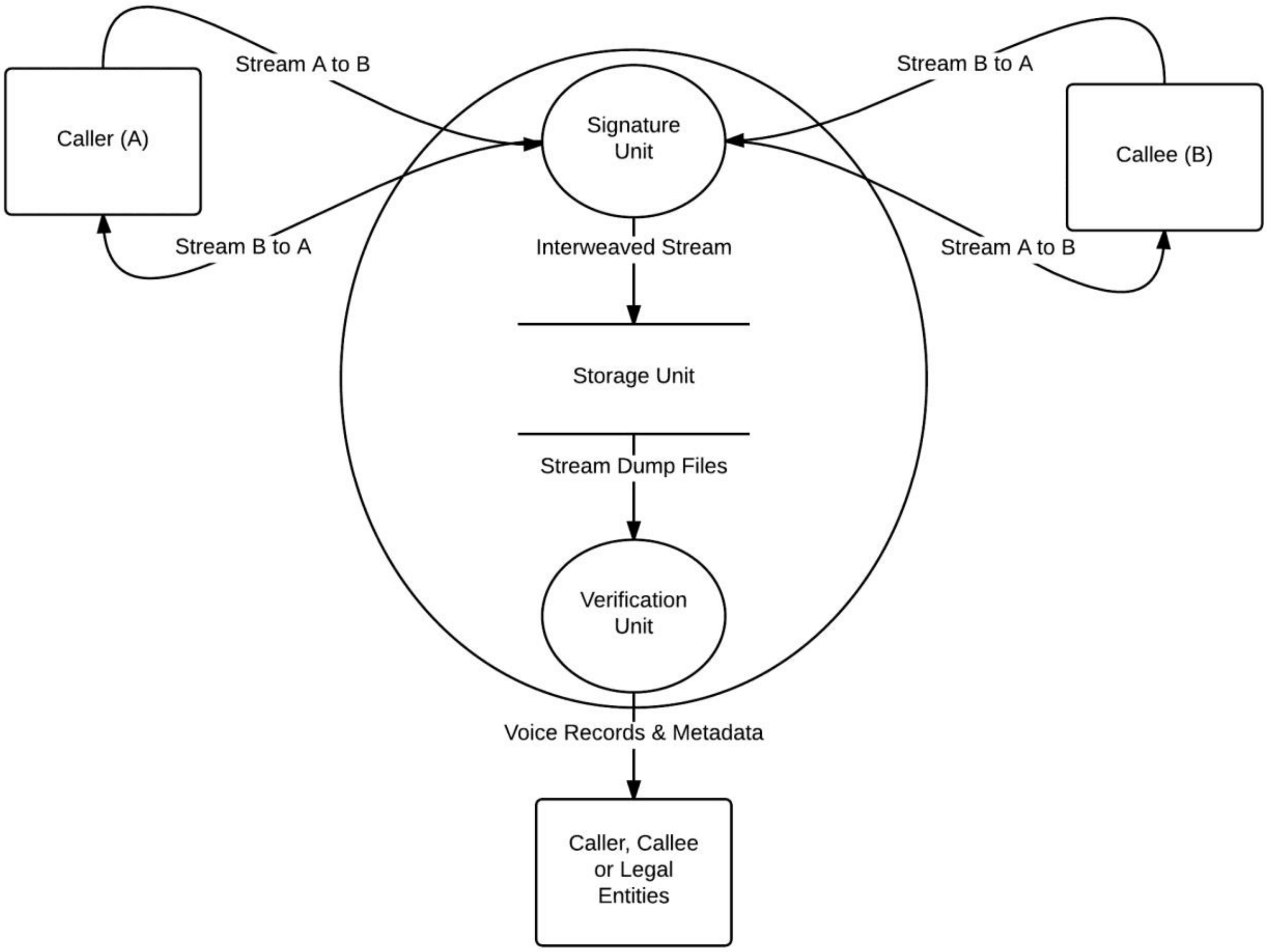

Figure 13: Data flow diagram that shows cycle of voice data within the system [f13].

Data flow diagram given in Figure 13 visualizes the cycle of the voice data, which is the user plane data of the mobile network, within the system. Signaling data and requests are excluded. The big ellipse in the center represents the Centralized Non-Repudiation Core, introduced in Section 3.8. Voice streams those originate from one party and delivered to the other, are untouched and kept as original. Only the interweaved (unified) stream, which is generated using copies of the original streams, is processed by the components of the system. So, parties shall not experience any degradation or distortion, except for a negligible delay caused by the copying process.

### 3.5.3. Used Interfaces

In LTE, the UE communicates with the BS (eNodeB) using LTE-Uu interface. Transport of the voice data from eNodeB to SGW is done over S1-U interface. SGW and PGW uses S5 and S8 interfaces. Details regarding these interfaces and their protocol stacks are shown in the Figure 8. This standard scheme is kept as it is in the system as it is located in the IMS core. In UMTS, the scheme is quite similar with the exception of SGSN relaying. Figure 14 shows the configuration in a layered composition for LTE and 15 for UMTS.

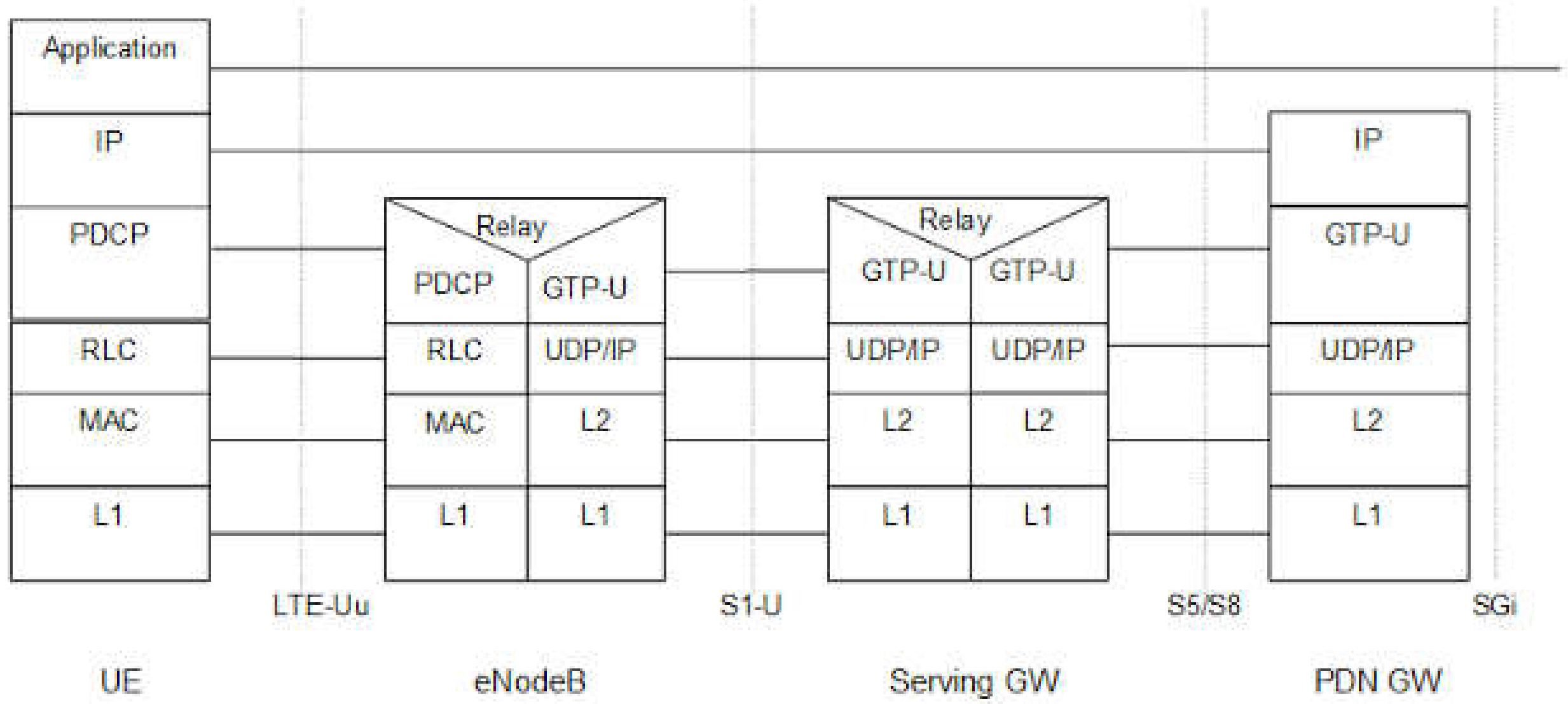

Figure 14: Communication protocols used in LTE RANs [f14].

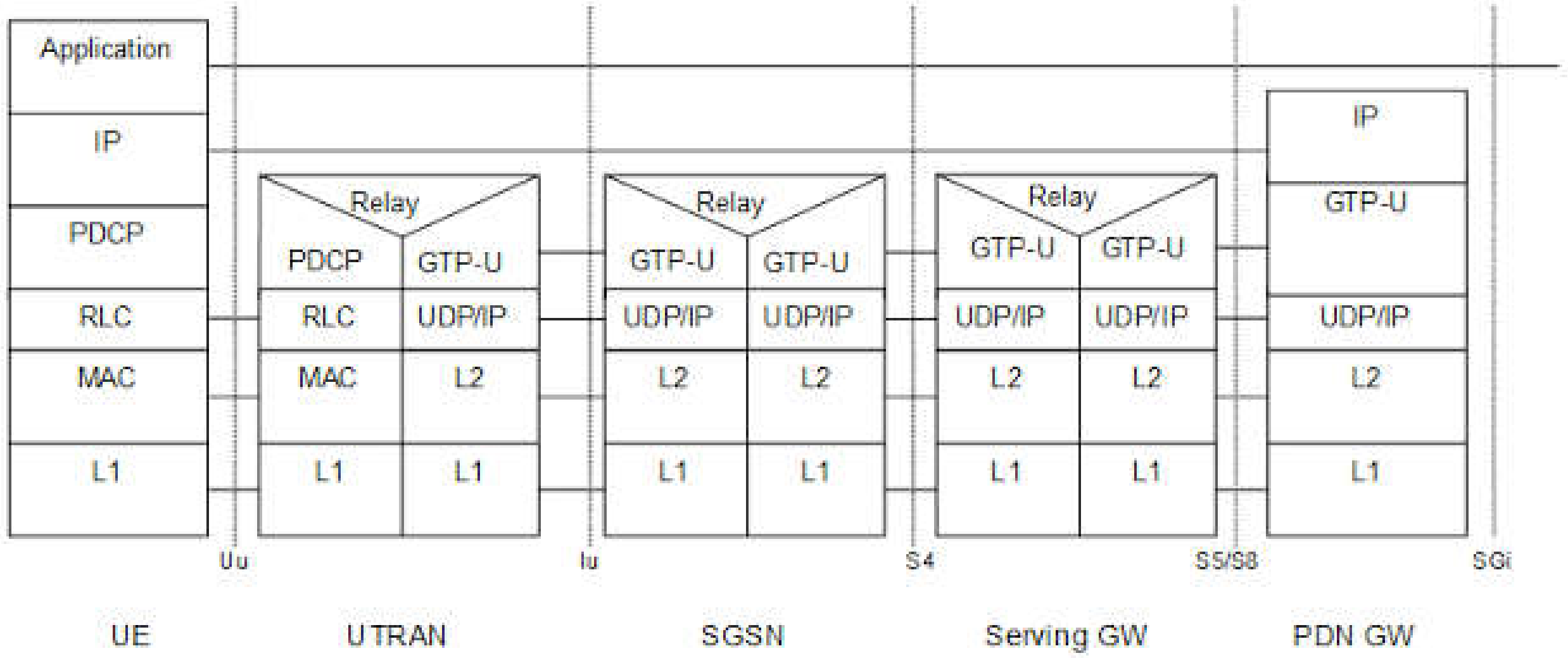

Figure 15: Communication protocols used in UMTS LANs [f15].

Actual voice data is carried in RTP packages, embedded in UDP datagrams, eventually inside IP packages. See Figure 1 in Chapter 1 for more details. Voice is coded using AMR (NB or WB) coding standard, which splits the stream into 20 ms long frames and codes with bitrates between 1.8 kbits/s to 23.85 kbits/s depending on the mode. Hence, creating intervals of the voice stream as introduced in Hett's work is already done in mobile networks. The coding mode, on the other hand, is chosen considering component capabilities, available bandwidth and QoS values by the MNO and it may even change dynamically during a call. But, the most common choice for UMTS and LTE is 12.65 kbits/s in practice during a clear call.

Transport of the voice packages to the IP backbone from the PGW is done by SGi interface. From this point on, voice data is on the IP backbone and is reachable by the MRF directly. MRF contains MRFP which is, in the system proposed, the signing unit responsible for electronically interleaving the incoming packages, signing them and forwarding them to the storage unit. In case of having the storage unit as a different device and a different network location, which is the recommended setup, there is no predefined interface nor file format for the communication between signing and storing units. This interface is defined within this work. Brief explanations about most of the related interfaces of IMS based cellular networks are given in the Appendix A.

For the communication between the signing and storing units, which involves the transfer of signed raw RTP packages, signature data, public keys and audio data, a TCP/IP connection should be preferred instead of UDP to prevent the packet loss by enabling retransmission of lost or corrupted packets, this is because signed voice records should be the same as genuine call data. So a basic RTP relaying scheme over UDP wouldn't be useful.

## 3.6. USSD/USSI Application Server

USSD/USSI Application Server is a key component for the system located outside of the IMS Core Network, but connected to IMS core and to the other sub-networks of the MNO over the USSD/USSI Gateways as explained in Section 2.7 and partially (for USSD) shown in Figure 9. As mentioned earlier, while Legacy GSM and UMTS natively supports USSD, LTE makes use of a simulation engine called USSD over IMS (USSI) that resembles the USSD functionality. In any case, since the application server is behind the mentioned gateways, role and configurations of the server doesn't vary. The same application server can serve both networks using different interfaces. USSD uses SS7/MAP protocols, which are CS based signaling protocols and USSI uses SIP which is the standard of IMS. Most modern USSI Gateways can handle both communication methods and can make conversions when necessary [21].

The application server shall have a special software to process signed call requests made by users and to invoke call-related units in the IMS core, namely HSS/HLR, PCRF and CSCF (which acts as another AS) using SIP to initiate a signed call. Sending notification confirmation messages to the parties is also a responsibility of the application server. Optionally a menu can be implemented, so that users can browse different actions including subscription to the service, PIN change or subscription cancellation.

## 3.7. User Credentials Database

The User Credentials Database is a collection of data about registered users of the proposed system. It is located in a dedicated physical database server in MNO's IP network. It can be a standalone DB server in the IMS Core Network. But for flexibility purposes, as another possible solution, it can also be located in the USSD Application Server introduced in Section 3.6. In any case, USSD application server shall access to this database to make user authorization requests and, depending on the implementation, even to make new user subscriptions or removals. The database is required mainly for three reasons. First, to prevent unauthorized usage of the system with spoofed devices. Second, to provide backwards compatibility when a user changes his credentials or other information. Last but not least, to transmit user data in fast and easy way between network elements of the MNO. So the system does not directly rely on the information that exist in the HSS.

Because of functional flexibility reasons, it is strongly recommended to implement this database in a separate server, which allows access from different interfaces. Because of the same reasons, XML is recommended to be used to store user credentials in a log style DB file.

Here is an example of a DB entry that contains user details.

```
<?xml version="1.0"?>
 <Users>
   <User>
       <SubscriberID>+491234567890</SubscriberID>
       <UniqueServiceID>000001</UniqueServiceID>
       <PIN>1234</PIN>
       <ServiceEnabled>1</ServiceEnabled>
       <RegDate>01.01.2015</RegDate>
       <FullName>Umut Can Çabuk</FullName>
       <Address>Fraunhofer SIT, Rhein Strasse 75</Address>
       <City>Darmstadt</City>
       <CommercialEntity>0</CommercialEntity>
   </User>
</Users>
```

Subscriber ID is the phone number of user's registered UICC (SIM card). Unique service ID is an optional ID, given to registered users of that system. PIN is the authentication code requested via a USSD code before a call is made. It can also be stored as a hash of the original PIN code to increase confidentiality, if provider cannot guarantee the security of the database depending on the implementation conditions. Service enabled field is a flag that indicates that user is registered and did not cancel his subscription. Name and address information are also required to meet legal obligations.

Some of the information, like full name and address can also be collected from the HSS/HLR if already exist, otherwise user should provide while subscribing. Commercial entity field indicates whether the user is a company/institution or an individual. This may affect types of extra services or billing plans. However this won't affect the proposed architecture, key functionalities nor schemes. Depending on the implementation conditions much more fields can be included. Except for the legal obligations, user information records are transparent to the system.

## 3.8. Centralized Non-Repudiation Core

Centralized Non-Repudiation Core is the combination of network components which provide non-repudiative operations like electronically signing the voice stream, storing the signed records and managing the user credentials. Core is located in the providers' IP network.

### 3.8.1. Signature Unit

The Signature Unit is typically an application that runs on a dedicated IP media server which has access to the user plane data on the IP backbone of the MNO. Fortunately, this role, namely an IP based media server that is able to reach voice data on the go, is already defined in the IMS framework. That role is called, as already mentioned in previous sections, MRF (See Figure 5). However a signing operation is not defined within this concept. So a special processing application on the media server is necessary to sign and forward incoming packages. To do that, the media server also needs a sufficient buffer memory to keep and process the streaming voice data.

MRFP, the processor segment of MRF, which is also responsible for audio processing operations like noise removal and echo cancellation, is able to make copies of the incoming IP or RTP packages. Controller part of MRF is MRFC, and is responsible for managing operations of MRFP depending on the received control messages from CSCF or from other IMS units.

Tasks of the signature unit:

- Creating signatures using users' credentials and timestamps.
- Generating copies of streaming VoIP packages to process.
- Signing voice data encapsulated in RTP packages.
- Forwarding signed voice data to the storage unit using IP communication.

### 3.8.2. Storage Unit

The Storage Unit (or server) is where the signed voice records are passively or actively stored as network traffic dumps and also as audio files separately. Additionally, signature data, public keys of signatures and metadata of the calls including time, duration and parties' identities are stored in this server in a database file. It is a network location that can be accessed over IP, which acts as an active file server with database capabilities and relevant applications. Storage unit could also be physically merged with the signature unit, namely MRF or MRFP, but because of the flexibility requirements, it has to be an IP based network location. Other than keeping voice data and relevant information, this property will allow other units like the verification unit, to reach and get the recorded data from the storage unit. And it may occasionally make it easier to get recordings from different signature units in a comprehensive and collaborative implementation mentioned in Chapter 4.

The Storage Unit must use a file system that is reliable and widely supported by different operating systems. FAT32 is not a very suitable solution since it cannot support files bigger than 4 GB, a voice call dump may exceed that value soon, if not today. NTFS is possible as it is widely supported, but its operation is not guaranteed to perform well under Linux and OS X. UDF, an open and vendor-neutral file system for data storage, appears as a more solid solution since it is supported by almost all recent operating system distributions, still the last choice is up to the providers [22].

Every information related to a signed call, excluding occasional intermediate files, recorded in the storage are considered as legally binding evidences. So, no intentional nor accidental change shall occur on these files. Only exception may be the cancellation made by caller using the final confirmation message over USSD (or SMS). In this case all information regarding that call will be removed from the server. Additionally, a newer signed agreement, either wet or electronic, between parties may declare the old as obsolete, but in this situation, old records won't be removed from the server unless a written request is delivered to the provider by both parties.

Main tasks of the storage unit are:

- Receiving and saving signed call streams from the signature unit, MRFP.
- Receiving and saving untouched but only decoded audio files of the streams in their original sound format.
- Saving signatures, public keys, user identities and call information in relation with the call dump files in an XML based database.
- Allow access to the relevant recorded voice data and metadata from other units in the network, when requested.

Storage unit itself doesn't process voice data. Hash generation and signing is done by signing unit, plus dump analyze and conversion is done by the verification unit. However in a future implementation, system can be extended with additional functionalities.

### 3.8.3. Verification Unit

The Verification Unit (or server) is the last (according to the data flow) functional piece of the non-repudiation core and is a web application server. It provides functions to verify signatures of signed calls and validate the call recordings, more importantly presents an HTTP/HTML based web interface to all users and legal entities to check, verify and download the signed voice records.

The users will be able to login to the web interface with their IMSI numbers and PIN codes or another pre-defined online transactions passwords. Hence they shall will be able to list and download all the signed and recorded conversations they have made using the service. Time, duration and ID information for the conversations shall also be presented as metadata to the users. In case of existence of legal restrictions on downloading the calls, an online player mechanism should be implemented to play the records on the web browser.

Major tasks of the verification unit can be listed as the following:

- Providing a web interface with login mechanism to users and legal entities that allows them to reach signed call records and their metadata.
- Requesting dumps of call records and their metadata from the storage server.
- List all signed calls made and their metadata to the user.
- Converting raw dump files received from the storage unit to actual sound files in AMR format (or exactly the one used during the conversation), remove audio watermarks if exist, and prepare the file to download or play.

For this last task, a special application based on WinPcap or libpcap that can analyze and process raw dump files should be used. Additionally, modules of Wireshark Network Analyzer application can also be utilized, which is an easier solution.

Actually, since storage unit is designed to store the audio files of the conversations as well as the signed dumps, if applicable, verification server can serve the recorded audio files directly to the users for them to play or download, instead of converting dumps to the audio files. But as a consequence, in this case there won't be an automated cryptographic signature verification. In fact, occasionally, signature verification may only be required when a disagreement between parties occurred or when there is a reasonable suspicion about the originality and authenticity of the call records. As long as parties have access to the records, a disagreement or a suspicion is usually a judicial matter. So, in such a situation experts can make the verification manually using the dump files.

## 3.9. Centralized Non-Repudiation Services

Centralized non-repudiation services are the essential operations executed by the components of the centralized non-repudiation core.

### 3.9.1. Packet Interweaving

In peer to peer systems, voice data is usually handled and processed as two different streams for each direction in both end points, namely each client processes incoming voice stream from the opposite client. However, in the centralized architecture, voice streams from both clients are gathered at a central point and can be handled as a single unified voice stream, which consists of speech of both parties. Therefore, the signature unit, which is a media server plays the role of the MRF in the IMS core, shall interweave incoming RTP packages from both directions in their orders of arrival before signing them. So that, an interweaved stream shall compose a record of a whole phone conversation. This operation is actually already being done in VoIP based teleconference calls as a standard procedure and the MRFP is supposed to be capable of doing this process as stated in definitive documents of 3GPP [23].

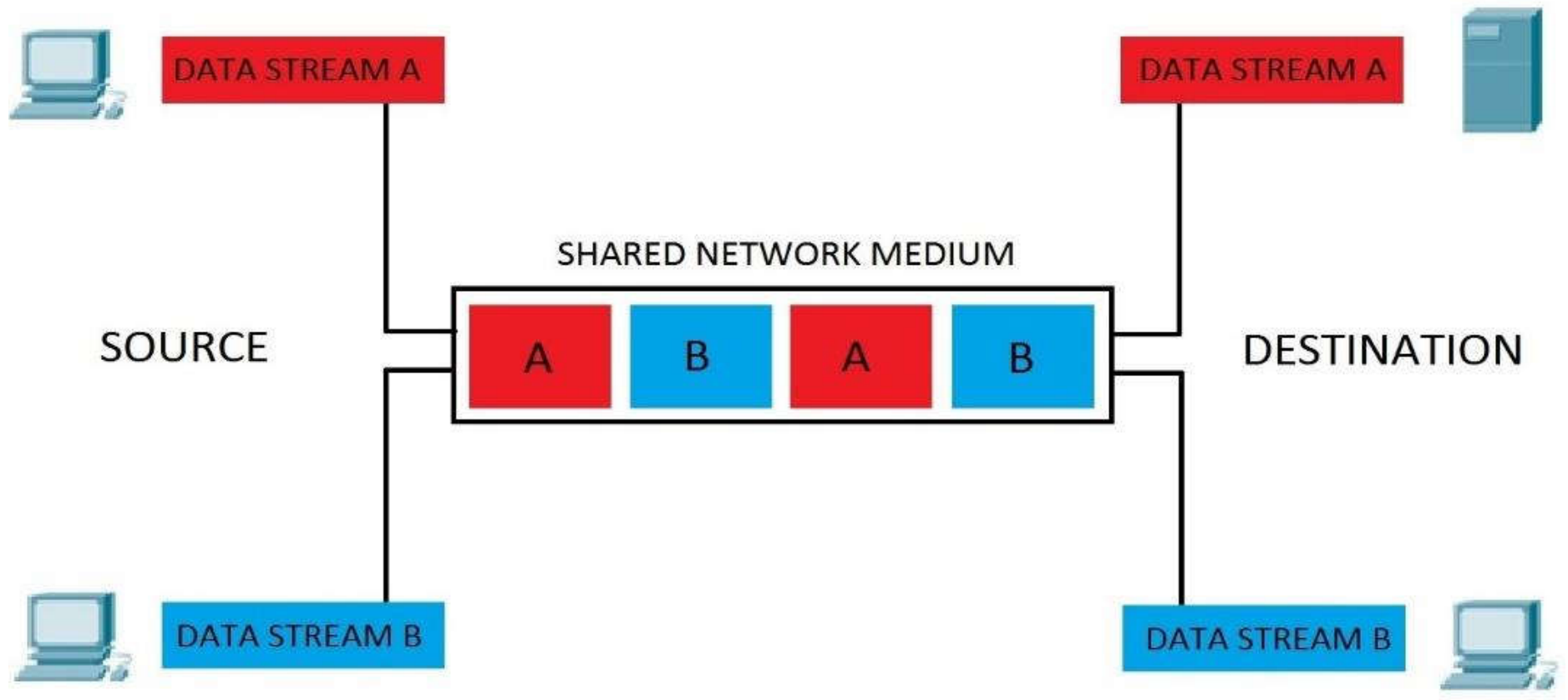

Figure 16: Packet interweaving example on a shared medium [f16].

Although it is not exactly the same case, Figure 16 gives an illuminative example of how an interweaved data stream should look like. In the MRF, streams are merged and captured as a single stream like the stream in the medium presented in that figure. After unification, signing process can begin.

Packet interweaving is not the only solution to prepare the data for signing and storing but it is the highly recommended one due to increased authenticity and integrity provided by signature chaining. Also this process will make a single stream as the output from two input streams, thus storage process may require less effort.

### 3.9.2. Audio Steganography

Audio steganography, which is also known as audio watermarking, is an optional step that takes place after interweaving and before signing. It basically means embedding another digital information into a voice data. Within the scope of this, work it can be useful to improve non-repudiation and integrity of the phone conversation. For this, identity information of the caller and callee, like phone numbers (MSIN), IMEI numbers or other information can be written to the RTP based voice packets in text or integer format using digital steganography techniques. By this way, in case of a disagreement between parties or subjects, it will be easier to find out the proof of origin of the recorded voice data in a further data mining analysis [24].

Usage of audio steganography in this system does not imply encryption of the watermark data as a must, watermarks can be in clear text format. Because a signing process will already take place after. As the data embedding method, least significant bit (LSB) coding in the RTP payload is proposed as a suitable solution. LSB coding algorithm replaces the least significant bit in some bytes of the carrier file to hide a sequence of bytes containing the watermark data. This will definitely cause a modification on the original audio file, but change of one bit per one byte is negligible for an audio file since human ear may not even detect the difference.

A revealed challenge is to have a voice stream with an unknown duration instead of a fixed size file as the carrier. So, it is complicated to spread out the watermark data over all packages. Instead, watermark data should be generated as small as possible and the same data should be embedded repeatedly in all received RTP packages. If watermark is not small enough, then sound could be distorted. Best size and other limitations for such a practice depend on used sound codecs and other communication methods, so it should be determined empirically.

### 3.9.3. Hash Generation

Hash generation is the first step of the two stepped signing process. Hash calculation is done by MRFP, for every RTP packet separately, by using secure cryptographic hash generation functions like MD5 or SHA1 or any other, the choice about the function is up to the provider, but it must have a fixed output size as a limitation. In fact, by the time of writing both MD5 and SHA1 are started to be considered as insecure and are deprecated from some applications. Newer and core complex algorithms like SHA2 (256) should be preferred in a real implementation.

Virtually, since the cellular networks have a “closed” or a “heavily regulated” network structure and unlike peer-to-peer VoIP communication done over internet, for instance Skype, used mediums and interfaces are also private. If we assume the level of security in the network as maximum, then storing only the calculated hashes without encrypting them would be enough to provide integrity and non-repudiation. However, this would be a very risky assumption in reality. Plus, that approach would limit the flexibility and scalability of the system as mentioned in Chapter 4.

### 3.9.4. Encryption of Hashes

Encryption of generated hashes is the second step of the two stepped signing process. As mentioned in section 3.2, this is a process that consists of asymmetric encryption of the calculated hashes of interweaved and queued RTP packages with the pre-calculated private key, and is done by MRF, in detail MRFP, during a call ideally in real-time. But alternatively, signing process can be done after the call is successfully terminated. In this case, recorded dumps can be used to make the operation of signing instead of the live stream. Even so, time gap between the call and signature process should be as small as possible, because that may cause a potential risk of the dumped voice data to be altered by attackers.

RSA, a PKI based encryption and signature algorithm, is recommended as the public key signature algorithm, since it is significantly popular in these type of applications and yet known to be secure. But other public key based algorithms like DSA, ECDSA, Schnorr, ElGamal etc. could be used, the choice is up to the provider. Running the signature creation algorithm is a task of signing unit which is the MRF of the IMS core.

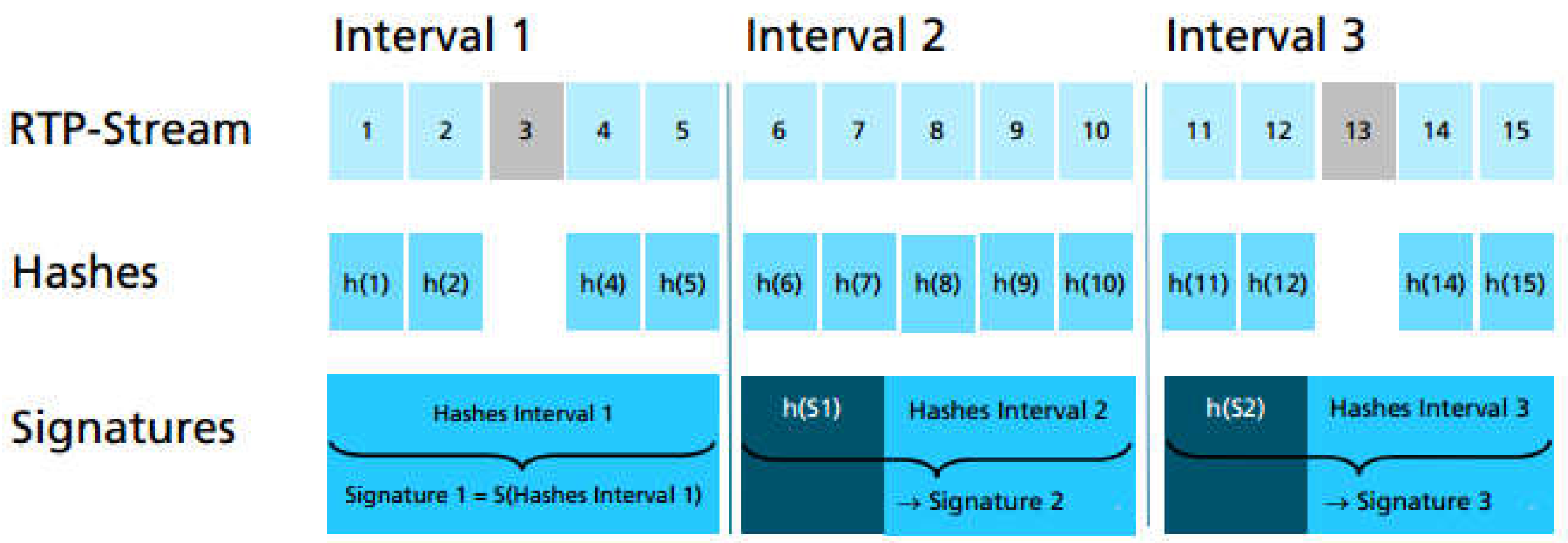

Figure 17: Hash generation and signature chaining on a stream with packet losses [f17].

For each signed conversation session, only one pair of private and public keys are generated. This is considered due to the wills of reducing required processing power, minimizing possible delays and above all keeping the output to be stored in the storage as small as possible. Each interval, videlicet a chunk of 5 consecutive packages, of the RTP stream is signed using the same private key during the same call. After the termination of the call, private key is no longer stored but disposed. Because, it must stay as a secret. If an attacker seizes the private key, authentication of the signature will be in question after that moment and that invalidates the whole signed call. Conversely, the public key is stored together with the signed stream in a database in the storage server and is used to verify the signature when necessary, for example, when user requested to play the call record later, using the web interface of the verification server.

The signature chaining mechanism explained in Figure 17. As shown, each signature also covers the hash of the one previous signature as well as hashes of the packages in that particular interval/chunk. If packet loss is occurred during the transmission of the stream, then hash of that lost packet is ignored and signing is made as normal. Signature scheme is explained in more detail in Hett's work [2].

### 3.9.5. Stream Forwarding

Stream forwarding is an automatic operation done by MRF, which stands for re-encapsulating the signed RTP packages into IP packages and sending them in their arrival order to the storage unit using the IP communication. Here, actual voice data and signature redundancy in RTP packages are not modified. But, unlike the previous communication schemes, RTP packages are transferred using a reliable TCP connection instead of instantaneous non-reliable UDP connection, which is mostly preferred in VoIP applications. Because, packet losses should be avoided and signed voice data as well as the signature data itself must be kept as original to sustain their legally binding property. That exigency implies generation of specific TCP headers containing information about the source, the destination and the data itself. Although UDP is already widespread, fortunately TCP is also natively supported by RTP. Alternatively RTMP, a real-time streaming protocol built on top of TCP, can be used to transfer the voice data.

IETF defines a method as a proposed standard to carry RTP streams using connection oriented protocols like TCP instead of UDP in RFC4571. Method implies re-framing of existing stream data and retransmission ways to prevent packet loss using packet and frame numbers. More details can be found in the document [25].

For security reasons, there should be no or minimum number of routing nodes between signing and storing units. However if the storing unit is located outside of the provider's network, due to legal obligations or collaborations between other providers, then this requirement may not be met. In such a case, all authorized parties must physically secure the connection. Alternatively, as a strong recommendation, TLS protocol can be used to digitally secure the connection between the signing and storing units.

### 3.9.6. Dump Storing

Dump storing signifies capturing and recording the dumps of an incoming TCP stream which contains RTP encapsulated signed voice data coming from the signing unit, namely MRFP, on a specific storage unit, which can also be a part of the MRF role. But this configuration is not recommended because of flexibility concerns. Capturing is done using passive packet sniffing or capturing, shortly known as pcap methodology implemented in libpcap and Winpcap libraries, which are used in popular network analyzing and packet sniffing applications like Wireshark, tcpdump, MS Network Analyzer etc. Captured raw dump shall be saved as a one-piece file in a format like .pcap, .pcapng, .cap, .dmp or similar. From this moment on, the dump file is a legally binding proof of the regarding phone conversation. Hence, no modifications nor retouches on the file are allowed. Although a lossless compression is possible, it is not recommended.

Among with the signed dump file itself, audio file of the call record (in .amr format or the one used originally) is also saved for practical purposes like instantly playing opportunity of the voice record when requested. Moreover, public key of the signature, call information including subscriber numbers, parties' identities, call date and time are recorded in an XML database file as well as the file path of the dump file. The database may contain much more information like ISDN ISUP report messages for multiple purposes. After each confirmed signed call the database is updated. If a signed call is later renounced by the caller, as he negatively replies the final confirmation request sent via USSD or SMS, then the recorded dump file and all database entries regarding the anterior call are to be deleted from the server.

## 3.10. Exception Handling

Because of the legal responsibilities, exceptional events regarding signed calls should be handled in a more constructive and informative way, for example than the entertainment services. Exceptions could be listed under three categories: Call anomalies, authorization problems and internal network errors.

### 3.10.1. Call Anomalies

Call anomalies is the general name for situations when voice conversations are disrupted because of technical problems which affects the quality or sustainability of the ongoing connection. Call anomalies for the signed calls can be analyzed under two cases: endurance and outage.

After a call is initiated, if network service becomes unavailable because of environmental effects or high-speed movement or internal faults etc. and if call is hung up, then both users shall have been informed about this unintended ending of the call in written form via a USSD notification and optionally with an SMS message. But ignoring all the conversation and automatically cancelling the signing process is not recommended. Because the important topics may already be discussed and call may already be about to end soon. So, in such a case cancelling the session causes parties to make the same conversation one more time, which is redundant. Hence, the best solution is to ask the caller if he still wants to sign and keep the record of the conversation even though it is ended unintendedly, via the same USSD menu message. In reply, caller has options to sign or ignore the call. This final confirmation message shall be sent in any case to provide opportunity for caller to quit the signing process.

On the other hand, if the signal level or Quality of Service (QoS) values in general goes low [9] but the call is not dropped during an ongoing call, no special action is needed other than the final confirmation message, which shall be sent to the initiator of the call at the end of any signed call. If parties cannot understand each other clearly because of bad call quality, then it is assumed that they will make the call again in any time. Call quality also does not have any effect on the signing process, because even though voice is not clear, RTP packages transferred on the line, will contain some digital voice data, which is enough to produces relevant hashes and signatures.

### 3.10.2. Authorization Issues

Since a call is initiated after authorization process is done, authorization problems may only occur prior to calls. Within this system, to increase certainty about users' identities, it is a necessity to implement additional verification layers introduced in Section 3.3. Standard recommendation is the PIN protection or similar text based alternatives. Users will be asked to enter their PIN codes via a USSD message right after they sent a request to make a signed call again with a USSD message, and same applies for the SMS solution. If a caller (or a hijacker) enters his PIN wrong then the call shall not be initiated, he shall be informed about the mistake and PIN shall be requested again in another USSD message. To prevent brute force attacks on the PIN codes, after three wrong attempts, signed call service shall be blocked to that caller in fact to his SIM card, for 30 minutes or a similar period. Preventions could be extended with a daily blockage after, in example, 6 wrong attempts, if decided as necessary. These steps are required to prohibit unbeknown people to abuse the system. No permanent blockage is considered, because in case of lost or stolen phone (SIM card) any user will presumably apply for a revocation to his MNO.

### 3.10.3. Internal Network Errors

During a signed call, any kind of device or interface fault in the core network may corrupt or interrupt at least one of the operations of the centralized non-repudiation core. However, unlike call anomalies these faults cannot usually be discovered by the talking parties and call goes on normally. Even though this problem statement is quite vague, the strategy to be followed is simple and clear. In case of such an error, principally signing and recording (dumping) processes should be cancelled, records made until that moment must be removed and users must be informed about the cancellation as a legal requirement, using USSD notifications and/or SMS messages. Optionally a sound alert like a "beep" tone can be implemented to warn parties about the cancellation of the process.

Internally, depending on the source of the fault, MRFP or Storage Server shall inform the PCRF, which is responsible for charging, through MRFC using the signaling channels. So that no charges will be applied for the signature process, but charges still may apply for the call depending on MNOs' policies. If an alert tone will be implemented, then generation of the tone is also a task of the MRFP.

## 3.11. Charging as of a VAS

The system proposed is advisable to be served as a Value Added Service (VAS) by MNOs or any other providers like Virtual MNOs. Thus, providers may apply fees to their users for usage of this service depending on their billing strategies. Substantially, there are some limitations on charging of such a system. This study comes up with two billing solutions: subscription and per usage.

Charging per usage refers to issuing a predefined fee per signed call. In principle, only the caller should pay the mentioned fee, but MNOs may apply different rules. Since a caller may cancel signing process right after the call is ended, by answering the final USSD query, fee shouldn't be issued until signature is confirmed by the caller. If flexible or negative balance feature is not supported by the MNO, then this necessity may cause exclusion of prepaid tariff users from the system. Because in parallel, regular fees for the call may also apply. This may occasionally consume a prepaid user's all credits and there may not be enough credits left for the fee of the signature service. So fee per usage method should exclude prepaid tariff owners if the MNO doesn't provide negative balance feature. Over and above, because of the legal aspects, the service should strictly be prohibited to ready-to-use style prepaid card owners, who did not sign an initial contract with the regarding MNO, which states identity and address information of the user. This plan is recommended to individuals and to who will rarely use the service.

Monthly or annual subscription is suggested as a more reliable billing method. In this case, callers and optionally callees must subscribe to the service for a month or a year and they pay a predefined subscription fee at the date of their invoice issuance. They have the option to stop or continue their subscription to the service whenever they want. Unlike the per usage method, this solution is also suitable for prepaid line owners. As long as they are subscribed, it won't matter if they have any more credits or not during their usage. Only, when they are out of credit at the renewal date, they won't be able to renew their subscription. But when they get enough credits, they will again be able to renew their subscription. This solution is recommended to businesses and to who will use the service frequently.

# 4. Strategies for Roaming and Interoperability Cases

The architecture and the operational schemes presented in Chapter 3 are designed to be valid only if the caller and the callee are using the same network provider, plus if they are in the coverage of their own home network. However, in practice, this design consideration may limit the usability of such a system dramatically. In case of roaming or usage of different possible providers, there must be special strategies developed. Since calls made in roaming situations become more and more popular, easier and cheaper by the time [26], especially between businesses, such a service should be extended so that it also covers roaming calls.

## 4.1. Usage in a Visiting Network

Usage in a visiting network generally refers to a roaming scenario, when at least one of the users is out of the radio coverage of his own MNO but have signal from a collaborative MNO, which offers similar IMS oriented IP based telephony services. By definition there is no statement on domestic or international availability, however because of commercial reasons, in industry, roaming usually refers to international collaborative access. From architectural point of view, there is no difference between international or domestic roaming. On the other hand, to support legal background about digital signatures in such a wide system, pertinent agreements must be made between countries in case of international roaming. EU Directives of European Parliament can be a good example as a framework to follow in this case [17]. These agreements can be various depending on the conditions, local laws and other agreements, so they are not specified in this work.

First of all, caller is the main actor in the whole service. He initiates a signed call session, he confirms the signing process and he is charged about the usage. Callee on the other hand, passively accepts or rejects an incoming signed call. This privilege comes not only because its spontaneity, but also a technical consideration about who shall provide the service. As long as a flat signature service which covers many clients in many MNOs in many countries is not implemented, it is assumed that the MNO of the caller party is responsible for providing infrastructure for such a service and meet its requirements, of course if they offer this system as a VAS to their clients.

### 4.1.1. Callee in a Visiting Network

If the caller is located in his home network but the callee is in a visited network, then this is the simplest roaming scenario, assuming both parties use the same MNO, which offers that service. In fact, this is, not literally but technically, the same case as the caller and the callee have different MNOs but geographically located in their own home networks. Because the service is already provided by the caller’s MNO. However such a use must rely on agreements between these MNOs.

In this case, since the actual voice data already passes through caller's home network's IMS core, the home network's MRF will have direct access to the raw voice data encapsulated in RTP packages and transported in IP packages. This access allows caller's home network to sign and store voice data within its own components without any extra implementation effort.

### 4.1.2. Caller in a Visiting Network

If the caller is in a visiting network, while the callee is in its home network, please note caller and callee use the same MNO in this case, then caller should virtually be moved to the home network's IMS core using IP tunneling features of the visiting network. Because streaming voice data eventually passes through home network's IMS core and MRF has access to the raw audio data. To achieve that functionality, two out of three IMS roaming models offered by related GSMA framework [27] can be utilized if they are or to be deployed by the collaborating MNOs.

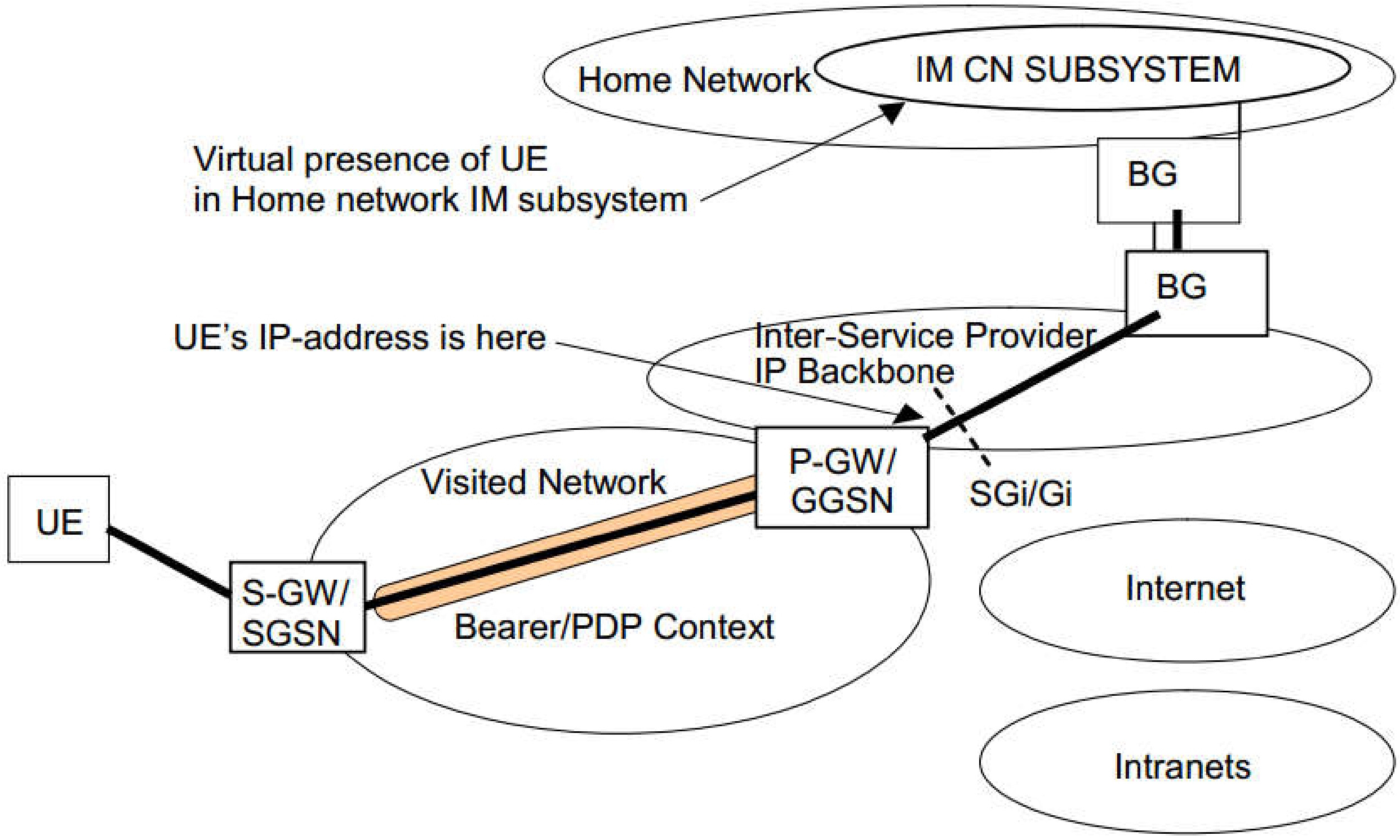

Figure 18: UE accessing IMS core services with P-GW/GGSN in the visited network [f18].

Figure 18 and 19 illustrates roaming configurations published by GSMA which virtually carries the visiting UE to its home network using data tunneling mechanisms. These two configurations allow VoIP data to pass through the IMS of home network. Thus, home MRF shall have access to the raw voice data. In the other configuration presented in the mentioned GSMA document, which is not shown in this section, tunneling ends in the visiting network's IMS and it does not allow user plane data to be transferred into home network.

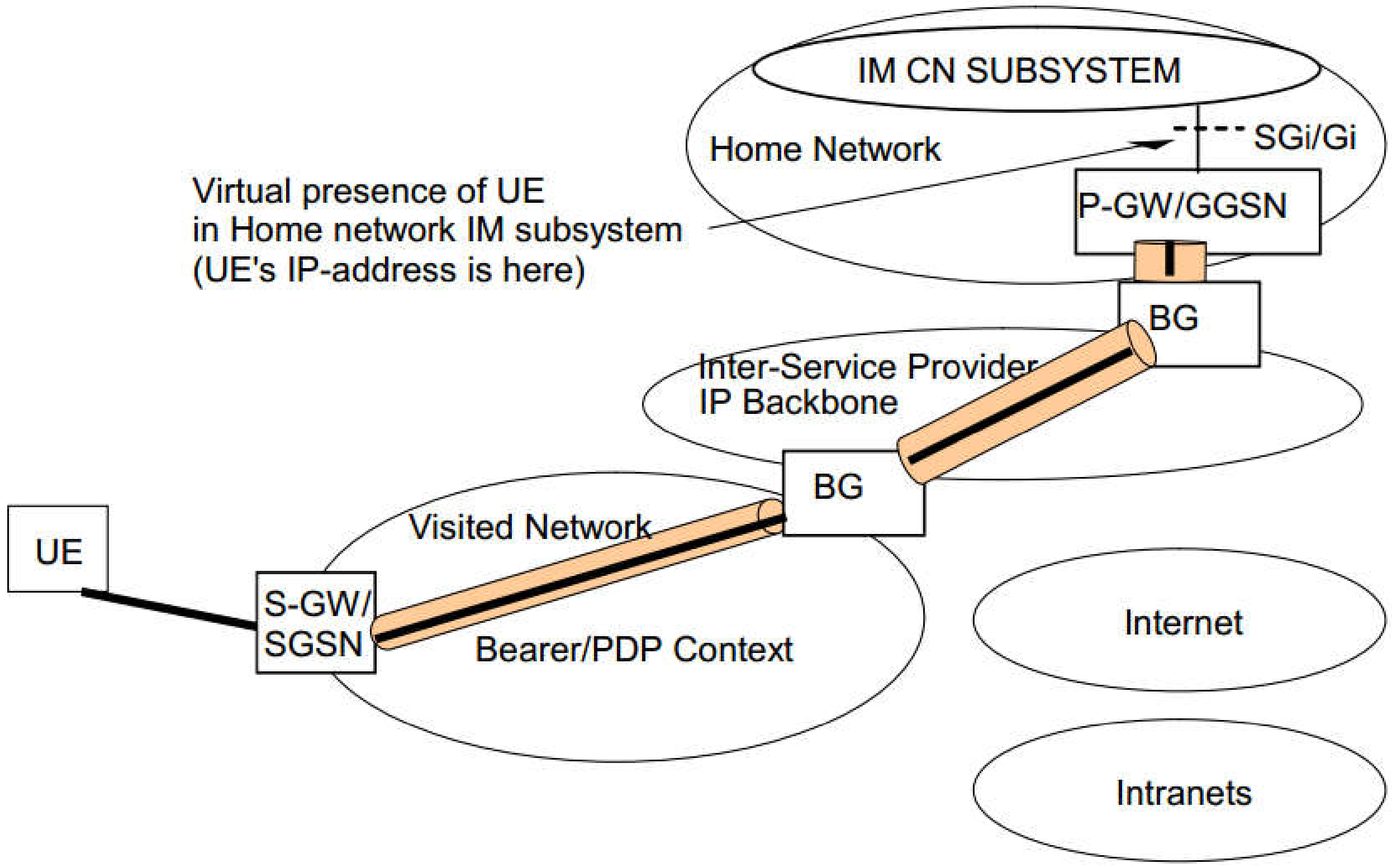

Figure 19: UE accessing IMS core services with P-GW/GGSN in the home network [f19].

Additionally, the USSD messaging service, as well as SMS, is always usable in a visiting network. Thus, a caller is always able to try to initiate a signed call using introduced call initiation methods in Section 3.4. In return, a home MNO is always able to inform the caller if the service is available or not, in the current place of the caller at that moment.

The call flow diagram given in Appendix B.2 also covers the roaming case and can be helpful to understand the communication schemes of a roaming call made using IMS backbones on both networks.

## 4.1.3. Both Parties in a Visited Network

If both parties are in a visited network, again assuming they use same MNOs, then naturally voice data may not pass through the home network's IMS at all. In this case, the strategy to be applied is similar to the one above. But here, UEs of the both parties shall be virtually moved to the home network by IP tunneling. So, configurations illustrated in Figures 16 and 17 shall be applied to both caller and callee. Hence, the home MRF shall gain access to the voice packages.

In such cases, custom roaming configurations can also be usable. However, any configuration that doesn't imply passing of voice packages through the home IMS cannot be used. The voice packages must eventually touch to the home MRF. A more comprehensive possible solution is presented in Chapter 6 as a future extension to this work.

## 4.2. Inclusion of Landline Telephony

It is assumed for both parties of that system to make use of mobile phones to use the service. But for companies and institutions with offices, branches or stores, relying strictly on mobile phones is not very suitable. In point of fact, landline phones are usually more preferable solutions. Hence, the system can also be extended to cover landline telephone networks under special circumstances using specific installations.

If a caller is to use a landline phone, then the only case which allows him to use the signed call service is to have an IMS integrated VoIP based call infrastructure on his home network. This setup is actually called as Next Generation Network (NGN) and is a wide research topic of Telecoms & Internet converged Services & Protocols for Advanced Networks (TISPAN), which is a standardization body of ETSI. This setup aims to move all landline (or fixed line) CS based conventional network services to the PS domain using simulation mechanisms [28]. See Figure 6 in Chapter 2 for the architecture of the TISPAN NGN. In this instance, landline operator is the provider. Additionally, the networks must support USSI and SMSoIP services. In fact, that setup refers to have a cable replacement for the RAN of cellular networks but to keep the core network structure with minor changes. Hence, if caller or callee or both are using a VoIP landline phone network which has IMS implemented, then configurations given in Section 4.1 are to be applied to provide the service depending on the situation.

Independently from the callee, if the caller uses a CS based PSTN line, then inherently it is not possible for him to use the signed call service and for the operator it is not possible to render the service to the PSTN users. A unique possible solution is to migrate the service to the callee's network only if he uses an IMS based network. However this option requires a different or an extended architectural design and is put out of this work. This scheme is partially explained in Section 6.2.

If caller uses an IMS based network, like LTE or UMTS, but the callee uses a PSTN line, then it is technically possible to run the service. Normally, most common approach is to make a plain CSFB and bind the call using CS domain by the LTE or UMTS network, during these type of calls. But in this case VoIP cannot be used. However to run the service, conversation must be merely carried over PS domain in the caller's home network. At this stage MGW in the PSTN interface of IMS comes into play. It is the IMS component responsible for conversion and transition of media data between different types of networks. MGW converts TDM boice data coming to and from the PSTN into RTP encapsulated voice data on the PS side. Thus, it will carry CS voice data to PS domain to be processed and of course, vice versa [28]. More details about this case can be found in the call flow diagram given in Appendix B.2.

## 4.3. Utilization of Circuit Switched Networks

Because of the continuing popularity of the usage of CS networks, the system may be extended to cover these networks. But due to the packet based nature of the service, this coverage will be very limited.

As stated above, callers cannot use a CS based network like GSM nor PSTN. This intrinsically eliminates the case which caller and callee both uses CS based networks. So the only case, in which connections from CS networks are supported is when the callee is using a CS based network. Independently from the callee, if caller is roaming, then configurations given in Section 4.1 apply.

The IMS framework, as partly mentioned before, has special components to handle calls made from or to clients using CS networks. Namely, BCGF, MGCF, SGW and MGW, also known as IM-MGW. Basically; BCGF is a SIP proxy processes requests for routing from S-CSCF when the S-CSCF has determined that the session cannot be routed using DNS. MGCF, is a SIP endpoint that controls protocol conversions and the resources in the MGW using H.248 interface. SGW is responsible for conveying the signaling data between the interconnected PS and CS based mobile networks. In example, IP based SCTP messages to TDM based MTP messages defined in SS7, and vice versa. IM-MGW is the only component which interacts with the actual voice data, it converts PCM/TDM voice of CS to the RTP voice of PS, and vice versa [28]. See Figure 6 in Chapter 2. Additionally, MGW can also change the audio codecs when two networks use different audio codecs like AMR, which is preferred in LTE and G.711 which is preferred in GSM.

When a call is made between a PS IMS caller and a CS callee; on the IMS side, MGW encapsulates the voice data coming from the CS network into RTP packages, then in UDP datagrams and ultimately in IP packages since it is directly connected to the IP backbone of the MNO. Hence, the voice data gathered by MGW can be sent to the MRF using IP connection. The MRF here runs the service functionality by copying, interweaving, signing and forwarding the incoming voice data.

# 5. Implementation

To prove that the proposed system can be feasibly built as a service to provide the introduced functionalities, a partial implementation is made qua a supportive work. In the implementation practice, some tasks of the Non-Repudiation Core, but mostly the Signature Unit located in the MRF application server, are conceptually realized by creating a dedicated standalone Java application (or a framework). Java language is especially chosen, since it is very popular in industrial applications [29], has good ability to interact with external applications and is platform independent. Another intention is to present an origin to the future researchers or developers by introducing tools, functions and parameters required. Moreover, ideas for a larger scale professional implementation are discussed.

## 5.1. Tools and Methods

In this section, equipment and technologies used to realize this implementation are presented with their appropriate settings as well as some suitable alternatives. First of all, Windows 7 is used as the development environment on the Intel i3 Dual Core test bed. That was not a strict choice; author's previous experience, hardware support for the network interfaces and existence of the related works were considered. Implementation would work fine on Linux and Mac OS X too, with few modifications. For the same reasons, Eclipse is chosen as the Java editor and compiler, most alternatives would work fine. Wireshark and its modules, which are used comprehensively, mandatory packet capturing libraries and custom Java classer introduced in detail, in next subsections.

As mentioned, this implementation proposes a program to demonstrate the tasks of the Signature Unit located in the MRF application server. Important information regarding realization of this demonstration, and furthermore, other means of IMS application development is explained in detail somewhere else [30].

### 5.1.1. Wireshark Network Analyzer

Wireshark® is an open-sourced (GNU GPL) popular network protocol analyzer program available for Windows, OS X and Linux environments, which allows computers to capture and browse all the existent network traffic using various network adapters and different mediums, with the power of WinPcap (for Windows) and libpcap (for Linux) libraries. See the Section 5.1.2 for more details about these libraries. The tool can be found in its official web page www.wireshark.com. The program can be used as a full standalone suit with its GUI or can also be used as partially with other, even custom, programs using its independent modules. In this work Wireshark version 12, which is the most current version by the time of writing, is used.

TShark, an independent executable module of Wireshark that captures all the data and signaling packets in network traffic, is necessarily used in this work. Except for having more advanced features like packet processing, it runs very similar to tcpdump and its Windows release Windump, which are pioneer but primitive packet capturing programs. It can filter packets to be dumped depending on their protocols, sizes, formats etc. So that unintended network traffic can be excluded from the dump files. Additionally, dumping can be scheduled using duration, packet count or dump size values. All preferences can be set using command line parameters. In this implementation, the following parameters are used or considered to be useful:

**-a** Used to define certain duration, size or file count values as automatic stop conditions for the capture process. Even though a signed call could be long or large in size, it is useful to set "big enough" limits to prevent malicious uses or misuses of the system and to save system resources.

**-b** Very similar to "-a", but instead of totally stopping the capture, it is used to create a cyclic ring buffer, which allows to limit number of dump files and their sizes or durations while the capture is still going on. In this case, for example, if the file number is limited to 5, then after 5 dump files are created, new captures are continued to be recorded to the very first file until the stop condition occurs or the stream finishes.

**-c** Used to define certain packet counts as automatic stop conditions. This is useful to divide the whole stream into chunks consist of 5 packets as explained in Section 3.2. Chunks are to be used in hash (signature) chaining process. A phone call stream may consist of thousands of chunks. Each chunk will be saved as a different file. After the hashing process is done, chunk files can be removed. It can even be set to 1, so that every dump file will contain only one packet, similar to the scheme proposed by Hett [2]. However, this parameter stops capturing at all, so program structure may be changed to restart capturing again.

**-f** Must be used to specify capture filters. Capture filters allow to dump only the intended packets and to ignore unintended types of traffic. Filtering can be done using source and destination IP addresses, port numbers and protocol types. It should be used to extract the actual voice data carried in RTP packages encapsulated with TCP from the signaling traffic between the MRFP and the storage server. For all available filters see the manual documentation of TShark [31].

**-i** Required to be used if multiple network interfaces exist on the computer/server and one or some of them should be used instead of all.

**-r** Should be used to capture the input from a recorded dump file instead of a live stream. Can be useful in intermediate servers that are to be used for various reasons.

**-R** Used to apply read filters. Works very similar to the capture filters. But used to read data from already recorded capture files instead of live streams.

**-w** Used to state the saving location of the output file, which is the filtered dump of the input stream.

**-W** Can be used to force the output to be saved in a specific file format other than the raw dumps. May be useful depending on the post-process operations in a future work. .cap, .pcap, .pcapng formats are supported, for more see the manual.

More detailed information about parameters and their usage can be found on the Tshark manual documentation [31].

Mergecap, another independent executable module of Wireshark, which is able to merge multiple capture files (dumps) into one. In this work, it is used to interweave RTP packet chunks to make a unified stream to be signed later. Here are some parameters of the Mergecap module.

**-a** Can be used to ignore the frame timestamps and to disregard the chronological order of the dumped packets. It allows writing all packets from the first input file followed by all packets from the second input file and so on for the next files. Not used in this implementation, but might be required occasionally.

**-F** Can be used to set the file type for the merged output file. If not used, default option will be used. For example, when inputs are .pcap, format, then the output will also be .pcap, if not used.

**-v** Can be used to print the number of merged packages for statistical purposes.

**-w** Should be used to state the saving location of the output file, which is the merged and unified dump of the multiple given input streams. Original input files are not removed.

For more detailed information, please refer to Mergecap's official documentation [32].

### 5.1.2. WinPcap and libpcap

WinPcap and its Linux equivalent libpcap, which is actually the original release, are the essential libraries required to capture and process the network traffic on the go, using host computer's specified network interfaces. Both libraries are distributed online under GNU GPL Licence.

In order to use functionalities related stream capturing mentioned within this work, WinPcap or libpcap must be installed to the system depending on the OS. Nevertheless, Wireshark already comes with an integrated WinPcap or libpcap release depending on the OS. In this work WinPcap 4.1.2 is used as a built-in module of the Wireshark Suite. However in case of a necessity to design a custom software without using modules of Wireshark etc., then these libraries must be used as a base. Detailed information about the usage of WinPcap can be found in the official documentation [33].

### 5.1.3. Java Classes

Within this work, to realize the functionalities in a modular way, three Java functionality classes are generated: Dump class, Hash class and Sign class. Additionally, not to provide a key functionality, but to support the program structure, two more classes, AVTransmit class and Main class are created.

As an important premise note; considering a large scale real implementation, in order to reduce the processing power requirements and minimize the delay, hashes are generated not from every single package but from chunks of several packages. This is slightly different than as explained in Chapter 3, but the overall logic of the scheme is totally the same.

Dump class (See 5.3.1) is responsible for capturing and saving packets of the received live streams coming from specified ports and network addresses (by the network staff or programmers in real case), which are actually the "signed" voice conversation between the caller and the callee, to a temporary directory as chunks that contain several packets. Furthermore, it is also responsible for interweaving the received stream chunks and unify them by this way. It runs Tshark and Mergecap modules with appropriate parameters to perform these.

Hash class (See 5.3.2) is responsible for generating cryptographic hashes of recorded and interweaved dump files. It uses pre-selected hashing algorithms and saves hash outputs in a separate file in regular text format. In fact, for every hash/signature chain link, after initial run, which involves creation of the output file, it appends newly calculated hashes at the end of the same output file, so that it makes a superhash that consists of multiple concatenated hashes. Other than preparing hashes to be signed later, this method has another role; to strengthen the non-repudiation concept. Because, it also allows signature chaining as well as chunk hashing. To achieve this, signature data should be written in the next hash output file by the Sign class in later phases. Thus, every superhash will contain the signature of one previous superhash, so the very powerful non-repudiation property explained in Chapter 3 can be set.

Sign class (See 5.3.3) is responsible for signing superhashes generated from hashes of several interweaved dump chunks. For the number of packets collected in a chunk, 5 is considered as an example throughout the theoretical parts of this work, but it is not a strict recommendation, and also please consider the premise notice given in this section. Signing is done using pre-selected signature algorithms. As a result of the signing process, a text based signature output file is created, which contains the generated signature, the public key used and the outcome of the very first verification of the signature. Additionally, the signature is also written to the next superhash file to chain the signatures as mentioned in the paragraph above. Though it is not implemented, the private key should also be saved temporarily to allow the usage of the same key for the ongoing signed call. After the call is terminated, the private key should be removed completely.

Main class is the backbone class of the framework and is responsible for initiating the program, running the functionalities by calling the classes mentioned above as much as required, arranging timings of the functions and finalizing the program. Main class is a short and flat class, but customized for this specific implementation and might be rebuilt for other implementations based on this project. It is not even necessary if operations of other classes are wanted to be performed manually on the test bed. Hence it is not included to the distribution of this implementation

AVTransmit class on the other hand, is responsible for transferring the recorded dumps and the signature files to the Storage Unit that is presumably located on the network, using RTP streams over IP. It is based on the AVTransmit2.java class, which is originally distributed by Oracle with some modifications. However, due to the time limitations of the study, it is not fully integrated to the framework.

## 5.2. Test Scenario

The test scenario, as mentioned in the very first paragraph of this chapter, covers a partial but a key role of the Non-Repudiation Core, and mostly MRF unit represents the Signature Unit. In the scenario, it is assumed that any two parties had already initiated and established a VoIP based signed call introduced in Chapter 3, so that two unidirectional RTP streams which carry voice data of both parties are arriving at MRFP again as an assumption.

The Java application written, as a partial MRFP simulator, first dumps the incoming RTP streams and records as two different file sets. But streams are not recorded as a whole recording, instead they are recorded as many little chunks, which contain 5 RTP packages each to allow chunk hashing and signature chaining operations in later phases. Then MRFP application performs the packet interweaving operation introduced in Section 3.9. Interweave operation is done using timestamps of RTP packages as the reference of their chronological orders. Right after, hashes of each chunk are calculated using one of the hash algorithms like SHA-1 algorithm. Algorithm is chosen as an example, different algorithms can be used too, without modifying the main structure of the code, but algorithm choice must be consistent for the whole process. For example 5, or any other predefined number of calculated hashes are then concatenated and saved in a text file to piece together a so called superhash. Additionally, if exists, one previous signature value is also appended later to that superhash file. This method provides signature chaining property which has a key role on the non-repudiation feature. Later on, signing process begins after hashing is done. Program handles the superhash file, signs it using a PKI based signature algorithm and generates the signature data as the output. In this implementation RSA and DSA algorithms are tested as examples. Together with created public key and the algorithm information, this signature data is saved to another text file. Nevertheless private key must not be stored anywhere. Furthermore, signature data is appended to the next superhash file as mentioned above. So, except for the capturing process which continues independently until the streams are finalized, the hashing and the signing processes end for a single chunk of RTP packets and starts again from scratch for the next consecutive chunks, until all the recorded chunks are processed. In this implementation, hashing and signing phases are initiated after the stream is ended and totally recorded. This does not cause any problems, in fact it may even provide better performance than real-time processing in a real implementation.

After all process explained here are done, the final task of the MRFP (Signature Unit) is sending dump files and generated signature files to the Storage Unit using one of live data transfer methods like RTP. Superhash file is an intermediate output and should not be sent. Since the MRFP is not intended to store files, after the transfer, all files can be cleaned up from the Signature Unit's memory. Functionalities given in this paragraph are not implemented.

## 5.3. Implementation Details

In this section, detailed information and design decisions regarding the composed java program/framework are explained. Some examples from the implemented code are also presented to objectify some of the concepts. Additionally, full code regarding the implementation is to be delivered as a Java project file, separated from this thesis document. First of all, it is good to state that Java 7 is used for the development project. Although no special extensions are installed, in future works, use of earlier versions may require modifications on the code or structure.

Besides, using VLC Player application (can be found in www.videolan.org), two live streams are initiated from the test bed computer to its local network as a broadcast with a destination address of 255.255.255.255 and ports 5002 and 5004 respectively. Streams are received again by the same computer using the written application. Use of loopback address, although it was possible too, is especially avoided to observe the capturing behaviors and to make the experiment more realistic.

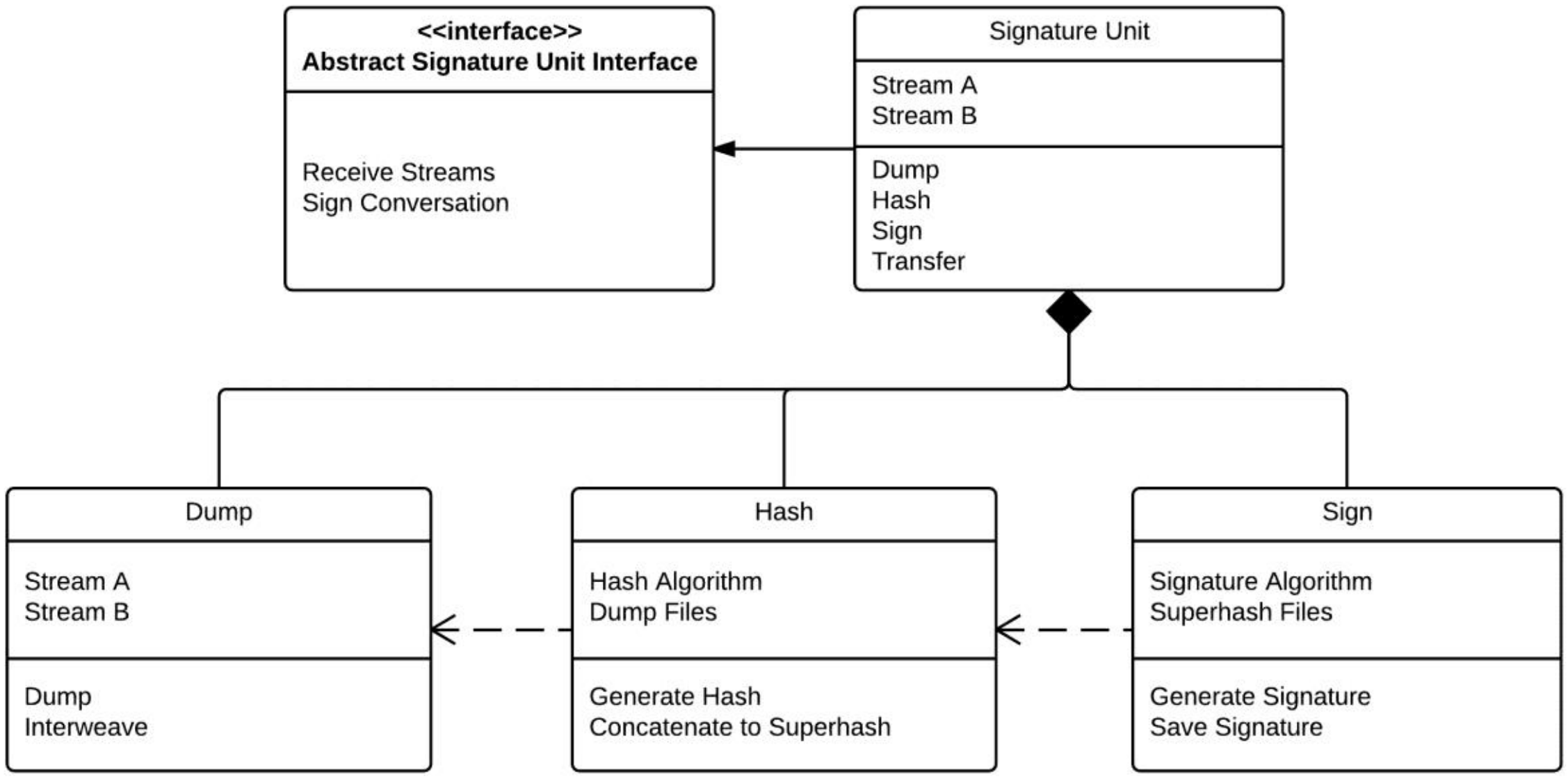

Figure 20: A Class analysis diagram of the functional structure of the implementation [f20].

Figure 20 represents the analysis version of the class diagram that belongs to the implementation made within this work. From the system-wide point of view, an abstract interface handles the data as intended. In detail, a signature unit runs, which is actually a superclass that is composed of other core classes to perform the defined functionalities. And as can be seen, core classes are dependent to each other to receive their corresponding inputs.

## 5.3.1. Dump Class

Named Dump.java. In the Dump class, two processes are created to run two instances of Tshark module with required parameters, which is used to capture the two incoming streams, as shown below. These two streams illustrate two RTP streams of the signed call shown in the Figure 13 in Chapter 3, A-to-B and B-to-A, respectively.

```
Process p1 = Runtime.getRuntime().exec("C:/Progra~2/Wireshark/tshark.exe -a duration:10
-b files:2 -b duration:5 -c:5 -f \“udp port 5002 and host 192.168.1.2\” -w
C:/Users/Umut/Desktop/stream1.pcap");
```

```
Process p2 = Runtime.getRuntime().exec("C:/Progra~2/Wireshark/tshark.exe -a duration:10
-b files:2 -b duration:5 -c:5 -f \“udp port 5004 and host 192.168.1.2\” -w
C:/Users/Umut/Desktop/stream2.pcap");
```

Both of the processes are identical except for their input sources which are two different streams. Because characteristics of the streams are the same and they will be saved in the same way.

Parameters in the code given above imply recording the UDP stream originating from network address 192.168.1.2 with port numbers 5002 and 5004 for the two processes. The IP address here is, presumably the IP address of the intermediate gateway located before the MRF. If streams are coming from different gateways, then these addresses must be different. Duration limitations are set for testing purposes only, they can be modified according to scenarios, in example, recording is made for 10 seconds in this implementation and each chunk is set to 5 packages. Additionally, as a prevention for problems that may be caused by the connectivity like packet losses, each chunk is limited to 5 seconds, by this way chunks will be saved even though the incoming packet count is not 5. For chunk duration, 3 and 2 seconds are also tested with no problems. The smaller the duration will cause the more the chunk count to be saved. Larger durations may harm the non-repudiation property. With reference to the premise note given at this section, to record single packets instead of chunks, as defined in Hett's scheme [2], packet count parameter c should be 1 instead. So that, every dump file would contain only one packet and hash generation would be made for every single packet.

There is no flawless way to filter solely the RTP packages and discarding all other types while capturing the network traffic. However, if the given host addresses and port numbers are predefined to transmit RTP streams only, which is a typical assumption for an industrial implementation, capturing the UDP traffic in the given way shall work fine. If traffic contains data regarding other protocols, then the following filter parameter written for the Wireshark GUI can be converted to TShark format using the TShark manual [31].

```
udp[1] & 1 != 1 && udp[3] & 1 != 1 && udp[8] & 0x80 == 0x80 && length < 250
```

This filter code basically attempts to recognize the header pattern of an RTP packet in incoming UDP datagrams. It may allow very few unrelated packets which meet these criteria to be captured, but it will include all related RTP packets definitely.

The code presented below is used to recognize all files in a single directory, which is the recording directory, using the name pattern carefully considered during the dumping phase and creates a file array which holds every chunk file in that directory. And paths of these files make a string array containing full file paths of the stream chunks in the directory using getPath() function. All chunks have a file name including the word "stream", a count number and reception date and time. Merging process itself is made by executing Mergecap module of Wireshark.

```
File dir = new File("C:/Users/Umut/Desktop/");

File [] files = dir.listFiles(new FilenameFilter() {

public boolean accept(File dir, String name) {

return name.startsWith("stream"); }});
```

Mergecap can merge many files at a time with a single line of parameter containing paths of all the files to be merged. Total number of recorded chunks to be merged is normally not known to the computer program and also is not predictable, because duration of a signed call cannot be estimated. So, the program follows the decided file name pattern "stream…" to select files to merge. So, the code fragment containing a while loop below creates a long parameter that contains paths of all the chunk files, for Mergecap to interweave at once. File paths are stored in a string array as mentioned and in each cycle of the loop, the program appends a new single chunk name to the already merged string of chunk names. The loop ends when the last chunk name is appended to the string.

```
int k = 0; int m = files.length; Process p=null;

String parameter = "-w C:/Users/Umut/Desktop/interweaved.pcap"; //Interweaved output
```

```
while(k<m-1){

       parameter = parameter+" "+ files[k].getPath()+" "+files[k+1].getPath();

       k=k+2;}
```

The code snippet below comes right after the one above and runs a process which executes Mergecap with the long parameter built by the code given above, that contains paths of all the files to be merged.

```
p = Runtime.getRuntime().exec("C:/Progra~2/Wireshark/mergecap.exe "+parameter);
p.waitFor();
```

By this way, the interweaved stream is created when the process terminates successfully.

### 5.3.2. Hash Class

Named Hash.java. In the Hash class, checksums or hashes of each chunk file are calculated via the function given below. Standard Java packages (i.e. java.security) support MD2, MD5 and SHA1 and SHA2 algorithms [34]. For others, additional libraries may be included. Checksum (hash) creation function is given below. It stores input files in a buffer and runs the hashing function defined in java.security library

```
public static byte[] createChecksum(String filename) throws  Exception {

       InputStream fis =  new FileInputStream(filename);

       byte[] buffer = new byte[1024];

       MessageDigest complete = MessageDigest.getInstance("SHA1");

       int numRead;

       do {

              numRead = fis.read(buffer);

              if (numRead > 0) {

                     complete.update(buffer, 0, numRead); }

       } while (numRead != -1);

       fis.close();

       return complete.digest(); }
```

The byte array created with this code below is later transcoded to a hexadecimal string using following function. Transcoding is done because hashes will be recorded in text format.

```
public static String getSHA1Checksum(String filename) throws Exception {

     byte[] b = createChecksum(filename);

     String result = "";

     for (int i=0; i < b.length; i++) {

       result += Integer.toString( ( b[i] & 0xff ) + 0x100, 16).substring(1);}

     return result; }
```

The code given below is used to append calculated hashes into one text file, which will then become the superhash file to be signed later by the Sign class. A superhash file contains 5 or another predefined number of hashes. After the number is reached, the file will be finalized and a new superhash file will be created. All superhash files except for the first one, shall also contain the signature of the previous superhash file as well as calculated hashes.

```
BufferedWriter outf = new BufferedWriter(new
FileWriter("C:/Users/Umut/Desktop/superhash.txt", true));

outf.write(getSHA1Checksum("C:/Users/Umut/Desktop/output.pcap") + "\n");

outf.close();
```

In this implementation consecutive hashes are written in consecutive lines. Adding CR+RF characters which provide the transition to a new line does not harm the functionality, but makes the file easier to read by humans.

### 5.3.3. Sign Class

Named Sign.java. The Sign class is the part of the program where superhash files that contain hashes of recorded chunks are signed. Calculation and verification of the signature is made using the functions given below. Input data is handled as blocks of 1024 bytes. Any other method or function could be used without harming the main structure of the program as long as they are able to handle custom sized text files as inputs.

```
private static byte[] sign(String datafile, PrivateKey prvKey, String sigAlg) throws
Exception {

    Signature sig = Signature.getInstance(sigAlg);

    sig.initSign(prvKey);

    FileInputStream fis = new FileInputStream(datafile);

    byte[] dataBytes = new byte[1024];

    int nread = fis.read(dataBytes);

    while (nread > 0) {

      sig.update(dataBytes, 0, nread);

      nread = fis.read(dataBytes); };

    return sig.sign(); }

private static boolean verify(String datafile, PublicKey pubKey, String sigAlg, byte[]
sigbytes) throws Exception {

    Signature sig = Signature.getInstance(sigAlg);

    sig.initVerify(pubKey);

    FileInputStream fis = new FileInputStream(datafile);

    byte[] dataBytes = new byte[1024];

    int nread = fis.read(dataBytes);
```

```
    while (nread > 0) {
      sig.update(dataBytes, 0, nread);
      nread = fis.read(dataBytes); };
    return sig.verify(sigbytes); }
```

Please note that, the signature algorithm here also applies its own hashing algorithm prior to signing operation. This done to match the actual data input to the suitable input format using padding schemes.

The following code portion is used to initiate the signature process in the main() function of Sign class.

```
KeyPairGenerator kpg = KeyPairGenerator.getInstance("RSA"); //"DSA" is also available
kpg.initialize(512); // 512 is the keysize.
KeyPair kp = kpg.generateKeyPair();
PublicKey pubk = kp.getPublic();
PrivateKey prvk = kp.getPrivate();
```

The code part below is exactly the same as the one given the Dump class, it reads all the stream files in the recording directory. This is later required to generating signatures from every single chunk file and to append them to the superhash file which already contains hashes of each chunk.

```
File dir = new File("C:/Users/Umut/Desktop/");
    File [] files = dir.listFiles(new FilenameFilter() {
        @Override
        public boolean accept(File dir, String name) {
            return name.startsWith("stream");       }});
```

The code block below calculates signatures from every single chunk file and appends them to the superhash file that already contains hashes. So that, the superhash file will contain hashes and signatures from every single chunk file. This operation is very important to achieve signature chaining which empowers the concept of non-repudiation.

```
String datafile = null; byte[] sigbytes = null;
while (n<m-1){
    BufferedWriter outf2 = new BufferedWriter(new
    FileWriter("C:/Users/Umut/Desktop/superhash.txt", true));
    datafile = files[n].getPath();
    sigbytes = sign(datafile, prvk, "SHA1withRSA"); //"SHAwithDSA", "MD5withRSA",
//"MD2withRSA" available
    outf2.write(new BigInteger(1, sigbytes).toString(16)+"\n");
    outf2.close();   n++;}
```

The following code fragment demonstrates how a superhash file, after appending processes are done, is ultimately handled and signed, plus how the output is saved. Hence, a chained or a cascaded signature scheme is obtained.

```
BufferedWriter outf = new BufferedWriter(new
FileWriter("C:/Users/Umut/Desktop/TEST/signatures.txt", true));

datafile = "C:/Users/Umut/Desktop/TEST/superhash.txt";

sigbytes = sign(datafile, prvk, "SHA1withRSA");

boolean result = verify(datafile, pubk, "SHA1withRSA", sigbytes);//verification

outf.write("Signature: "+new BigInteger(1, sigbytes).toString(16)+"\nFirstCheck:
"+result+"\nPublicKey: "+pubk+"\n");           outf.close();
```

As can be seen, the public key and the result of the initial check is also saved to the output file as well as the signature of the superhash itself. The public key is used to verify the signature later by automatic programs located in the Verification Unit or manually by the expert personnel. The initial verification is made to prove that the signing process is successful and there is no malicious activity in the Signature Unit.

## 5.4. Results and Evaluation

Within this implementation; key roles of the Signature Unit introduced in Chapter 3 are realized partially and in a simplified way. On the other hand, functionality, consistency and usability factors are carefully considered. So, here by this implementation, it is proven that any future industrial and/or commercial implementation of the proposed system based on this thesis study can be done and can be done using publicly available tools, libraries and techniques.

Namely, dumping RTP voice streams in chunks, chronologically interweaving separated streams, generating and hashes of recorded dumps, chaining hashes/signatures, creating signatures for calculated hashes, verifying signatures and producing output documents are the functionalities done in this implementation as tasks of an MRF application server which has a role of the Signature Unit.

Required functionalities and methodologies to realize such a system are presented in detail also with some alternatives. Likewise, how the MRF application server should be programmed is explained comprehensively. Furthermore, tasks of the MRFP (processor part of MRF) are presented in a distributed point of view. To achieve modularity, functionalities are tried to be separated in different classes and the task of controlling operations of these classes is kept out of the functional classes and included to the main class which acts as a backbone without any other special function. In fact, this controlling task can be moved to the MRFC (controller part of MRF) unit in a real implementation depending on the design considerations.

Prior to running the application, in order to demonstrate the voice conversation, two mp3 (MPEG-TS Layer 3, 320 kbps, 8.2 and 8.7 MB respectively) audio files are streamed over RTP using two different ports, to be captured via VLC Player. Then the application is run with corresponding filtering parameters, and then several output files are created as expected results. RTP chunk files from two streams, the interweaved stream of all chunks, a text based superhash file contains hashes generated from these chunks and a text based signature file contains signature data created from the superhash file, are found respectively.

Recorded chunk files are analyzed with Wireshark GUI version and it is seen that all of the chunks contain solely the UDP datagrams captured from the live stream which use preset ports. When these UDP packages are decoded as RTP using Wireshark GUI, MPEG-TS Layer 3 audio data was recognized in the packages. It was also possible to replay the audio data embedded in these packages. The same control is also made for the interweaved stream file. It was seen that the unified stream file contains all the RTP packages recorded in all of the chunk files. It was possible to replay two different audio data arrived from two different streams separately. Additionally some metadata like audio coding bitrate etc. was also available.

Generated superhash file is analyzed using WordPad. Any text editor with CR+LF character support would work fine, if not, all hashes would be displayed in the same line, this can make manual analysis harder. It was seen that the file was containing generated hashes from all of the recorded chunk files. Additionally, there were also signature data appended to hashes, which come from the signature chaining practice as expected. A hash output for 3 chunks (each containing 2 seconds of RTP stream), excluding signatures, will look like the following:

```
3ca03a9587b0e5d0a32ae89a1218942f21f5b639

8835593f9d912d96c24254352f10cbc274345bc2

511458a935b5864c641c7ffc71fe18b6f150ee6f
```

The signature file, which contains the signature of the superhash file, is also analyzed with the same text editor. A signature data, result of the initial verification, used algorithm, public exponent and the modulus which are parts of the public key, were listed in the file, as can be foreseen. A signature file will contain data that looks like the following:

```
Signature:
1913e068d0eb49c3fbcc6b63aa7c8a4ab60190da2114e58e81a62ba03f1882132284cacf41d6ca29e37269b
94d6d92c6366813f88f7378ca971003322b67df96

FirstCheck: true

PublicKey: Sun RSA public key, 512 bits

  modulus:
9417222498354992951390441550221290042461691467768319332379555451147406658286939170775775
56220787029077633511516817904469564131155747839050289746512524222291

  public exponent: 65537
```

Moreover, a basic performance analysis is made to question the practical feasibility of the implementation. To do that, runtime durations of some of the operations made by defined classes are measured using built-in Java time functions (java.util.Date). The time commands are wrapped around the code to be measured.

```
long lStartTime = new Date().getTime(); // start the timer

{…OPERATIONS TO BE MEASURED…}
```

```
long lEndTime = new Date().getTime(); // end the timer

long difference = lEndTime - lStartTime; // check the difference
```

Using the shown method, the following measurements are done, while two streams which carry the mentioned mp3 audios as payloads were dumped for 10 seconds. All experiments were repeated ten times.

| | |
|---|---|
| Merging ten (92 ± 18 KB) dump chunks: | 23 ± 7 ms (Chunk sizes shall affect the process time) |
| Hash creation for each (one) chunk: | 5 ± 3 ms (Except for the first run, it was ca. 60 ms) |
| Chained signature creation for one superhash: | 164 ± 35 ms |
| Copying data to buffered reader: | 1 ± 0.5 ms |

Apart from the others, the last measurement represents the real delay on transferring the voice data of the ongoing call in the core network of a MNO that is actually caused by the signed call service. If the call quality and user satisfaction statistics on the voice delay given in [9] are considered, then the delay caused by the signed call service is very small and can be accepted as negligible. The other delays will not affect the ongoing call, but related to the internal operations only. They have to be kept in a range that will not harm the real-time property of the service. More detailed performance studies should be made as explained in Section 6.4 as a future work.

All in all, as the evaluation of the practical work, experiments made using this implementation proved that, the Signature Unit functionality located in the MRF application server would be (partially, merely for the Signature Unit role itself) sufficiently successful to satisfy the needs and to fulfill the requirements determined by the goals of this study given in the Section 1.4. Especially the first three goals are virtually achieved with the results gathered from this setup. Plus, the outcomes of the experiments are also compatible with the use cases given in the Section 1.5. According to the outputs, the written application is performing the intended operations to realize a key unit of this signed call service, namely the Signature Unit. Besides, they also show that a professional industrial implementation of this system is possible with a reasonable effort, making it feasible in terms of software and system architecture, excluding hardware and peripherals.

# 6. Extensions and Future Work

Even though the work aims to provide a full solution to the non-repudiation problem mentioned, time and workforce limitations of the thesis project naturally constricted the coverage and depth of it. Besides, since the concept is quite new and not comprehensively studied so far, methodologies introduced within this research eventually unveil new areas to research deeper for considering detailed design decisions. In this chapter, a bunch of considerably high priority further work are listed and explained to guide future researchers and development engineers.

## 6.1. Advanced User Verification Techniques

Due to the time limitations given above, to keep simplicity and to focus more on the system architecture, in this work PIN codes are chosen as the second layer user verification technique. For real, PIN code verification is a very feasible method for such a system. However, as already mentioned earlier in Section 3.3, it is not very secure. Naturally, any technique that relies on use of PIN codes is clearly vulnerable to brute force attacks. Additionally users may easily forget their codes or be confused, in such a case renewal of the code can be quite hard, because the new code must be shared with the user using a (abstractly) secure channel. Hence, providers and/or legal entities may require use of more advanced techniques together with or instead of PIN codes.

As a future extension, use of secure tokens like electronic ID cards and USB dongles etc. can be researched and implemented. They may contain more personal information and in a cryptographic fashion. This option is initially left out of this work because of one of the main goals, which forbids use of additional software or hardware on the user side, to increase the usability and reliability of the system. So, a study to find methods to utilize physical tokens as verification method without vitiating the usability and universality of the system can be made. Notwithstanding, although physical tokens eliminate the risk of forgetting authentication information, they are prone to be stolen or got lost. So use of PIN codes should not be ceased, but offered as a secondary method instead. Alternatively both methods can be used together, but effectiveness of this case is questionable.

Use of biometric data arguably leads to the ultimate user verification methods. But unlike physical tokens, biometric information based verification systems may require radical functional changes or additions in the core network. For example, a voice recognition mechanism would require special applications on MRF or in dedicated authentication servers. Additionally, these methods may require different call initiation techniques. Although they might be seen as hard, expensive or redundant to invest, feasibility studies and possible architecture assemblies that rely on these methods deserve further research.

## 6.2. Complete Integration of CS Calls

As mentioned earlier, the conversation signing system introduced in this thesis requires voice calls to be made over packet data infrastructure as a matter of course. Voice data and the communication must be done using IP, RTP and SIP protocols, which are not supported by legacy circuit switched networks. Despite this fact, some interworking possibilities with CS based networks are given in Chapter 4. However, an extensive coverage for all CS based call combinations is not provided since it is seen as a different focus then the non-repudiation problem, which is primarily intended to be addressed in this work.

Be that as it may, since most of the mobile phone calls nowadays are still made over CS infrastructure, it could be very good to develop a consistent solution for bringing support to the CS calls. When a call is initiated by a UE utilizing CS domain, the most common strategy is to make a simple CSFB. But, instead a packetizing conversion can be made and voice data can be moved to the PS domain just to run the core non-repudiation functionality. This setup may require architectural modifications and is a subject to further research. One possible way to follow is trying to migrate the core non-repudiation functionality in the proposed architecture out of the MNOs networks and to rebuild it as an external 3rd party network which can offer the service to multiple MNOs like a cloud based operation scheme considering legal obligations.

Please note, the term CS calls contains incoming and outgoing calls from legacy GSM, UMTS (in CSFB case) and landline PSTN networks.

## 6.3. External Provision of the Signed Call Service

As a major improvement in the further development process, the system and the service introduced in this work can be moved from MNOs private IP networks to the public IP network, namely the Internet. Thus it will somehow be a cloud service open for multiple providers. This can be done via setting a special dedicated IMS core network on the internet by private companies or government institutions. So that, many MNOs, landline PSTN operators, ISPs that offer VoIP services as well as private IP PBX systems will be able to use the service without implementing their own infrastructure but using the shared cloud-style infrastructure. This architecture separates MNOs from the provider role and they become customers of these providers just like ISPs and private VoIP systems.

That revolutionary progress, in the big picture, will dramatically reduce costs to build such a functionality, since it removes the necessity for every telephony operator to build their own infrastructure. Moreover, roaming between different cellular networks and interoperability between different types of networks will be a less complex problems and to be addressed in the new emigrated architecture, not in every operator's networks. Additionally, legally induced tracings and regulations may be done much easier.

In such a case, call initiation methods, packet data routing solutions, CS-to-PS conversions, billing methods both for operators and end users, international usage and legal obligations must be researched again in a more centralized, modular and flexible way of thinking. Additionally, since the format of the voice data, like sound codecs and transfer protocols or user information style may be different in many different networks, a new mechanism to create and store signatures may be required.

## 6.4. Performance Studies

In such large scale networks with presumably thousands of potential users, especially if they are made different types of sub-networks which utilize different mediums, realistic performance analyses can only be done using prototype implementations. Estimations and simulations may be misleading and even infeasible to prepare, yet at the time of writing there is no comprehensive simulation tool available to public which contains all parts of an LTE network as a whole system.

Such an analysis should typically contain statistical data about transport delays of the voice data in different parts of the network, data write and read speeds in the server side while multiple calls are ongoing, memory usage in the packet data gateways and application servers during calls and processor power requirements on MRF role devices as well as other QoS values. Additionally, values from the user side, representing the user experience should also be measured to consider usability and user satisfaction level of the system.

## 6.5. Speech Recognition and Text Conversion

An extension to this work as a future study, could be automatic transcription of the speeches of both parties during a signed call and generating a text output from the conversation. Such an extension would increase the usability and preferability of the introduced service in industry, since provided text outputs can be used as legally binding documents by parties much easier than voice records.

This extension can be implemented by integrating an advanced human speech recognition algorithm to the system. In detail, an application server that has the MRF role is theoretically able to do such a speech recognition and text conversion, using the incoming voice streams. Afterwards, these text output can be stored in the storage unit and can be reachable when requested just like voice records. However, implementers must be sure about the reliability of the algorithm beforehand. No recognition mistakes can be tolerated within this system.

# 7. Conclusions

In this thesis work, a system architecture or a framework as an extensive subsystem of the existing IP based modern cellular networks to provide a non-repudiative voice call service with all essential methods and functionalities, is researched, designed, partially implemented and presented with outcomes. Meanwhile, main concerns which shape the structure of the work are compatibility with current systems, adherence to legal aspects, widespread usability and flexibility for the future expansions.

With reference to the use case scenario given in the Chapter 1; by using a real implementation of the service introduced in this work, a bank customer, who wants to make an official transaction like opening a bank account, which normally requires wet signatures, will be able to do this only by telling his intention viva voce to the bank officials on the other side of the line without the need of going to the office and sign related documents, instead voice conversation will be recorded and signed electronically by the service provider, making it a legally binding proof that is reachable when needed.

Regarding the main goals of the study, these statements can be made as outcomes of this work:

1. Non-repudiation concept is brought to the voice calls made over mobile (cellular) networks.
2. Solutions for digitally signing and storing the voice data from an on-going call in real-time and to track this recorded copy of the call afterwards, are introduced.
3. A centralized approach is taken into account.
4. Interactions and interfaces that provide ease of use and widespread support are recommended.
5. Compatibility with current LTE and UMTS systems are achieved via IMS using minimum possible customization. Solutions for including other means of networks are also studied.
6. Keeping the architecture as simple in order to make the system feasible and realistic for investors and to prevent conflicts with other current systems or services.
7. System is especially built in terms of, dependent but standalone, modules. Additionally many design alternatives and possible future works are proposed.
8. During the design of the functionalities to be served, legal requirements are accepted as guidelines. Ways are recommended to process and store sensitive user data. More, signed voice records are (hypothetically) allowed to be used as legally binding statements in case of an industrial realization.
9. Techniques are presented for providers to serve the functionalities of this system as a value added service in case of a commercial realization.

Moreover, a conceptual implementation of the proposed system, consists of writing and running a simulative Java program to test some of the key functionalities like signing the voice data, is made. That implementation showed that, the proposed theory is feasible and can be realized consistently.

# Glossary

**3GPP** 3rd Generation Partnership Project; An organization aims to develop new standards about mobile broadband telecommunication systems.

**AMR** Adaptive Multi Rate; An audio codec used in 3G and 4G mobile telephony networks.

**AS** Application Server; A server type which is dedicated to run specific applications.

**BG** Border Gateway; A network gateway, which stands between different networks of same type.

**CS** Circuit Switching; A communication methodology, relies on continual connection between peers.

**CSCF** Call Session Control Function; The role for devices and applications to manage call sessions in IMS.

**CSFB** Circuit Switched Fallback; The backwards compatibility mode, makes use of circuit switching in primarily PS oriented networks.

**DB** Database, A structured collection of information in various categories.

**EPC** Evolved Packet Core; The core network in 4G cellular networks.

**EPS** Evolved Packet System; The generic name for IP based 4G networks.

**GSMA** GSM Association; An organization aims to develop new standards about GSM based mobile telecommunication systems.

**GSM** Global System for Mobile Communications; A standard for digital mobile communications, yet called 2G.

**HLR** Home Location Register; A central component which keeps general information about registered users in cellular networks.

**HSPA** High Speed Packet Access; A pair of protocols that improves performance of 3G mobile communication networks.

**HSS** Home Subscriber Server; The HLR replacement with extended features in LTE/EPS core networks.

**I-CSCF** Interrogating Call Session Control Function; The role for devices and applications to …

**IETF** Internet Engineering Taskforce; An organization aims to develop new standards about the internet and related technologies.

**ISDN** Integrated Services Digital Network; A set of communication standards for simultaneous digital transmission of voice, video, data, and other network services over PSTN.

**ISIM** IP Multimedia Services Identity Module; An improved SIM card application for IMS based mobile telephony networks.

**ISP** Internet Service Provider; The generic name for landline internet infrastructure providers.

**ITU** International Telecommunications Union; An organization aims to develop new standards about telecommunication systems.

**IMS** IP Multimedia Core Network Subsystem; The core network framework for contemporary IP based mobile networks with extensive multimedia processing capabilities.

**KSI** Keyless Signature Infrastructure; An electronic signature mechanism, that can be used without private keys, except signer authentication.

**MAF** Multi-Factor User Authentication; A concept, refers to apply multiple authentication methods sequentially to authenticate users.

**MGW** Media Gateway; The gateway that converts and transfers the media data between different types of networks like CS and PS.

**MME** Mobility Management Entity; The control node in the LTE access network.

**MNO** Mobile Network Operator; General name for service providers in the cellular networks sector.

**MRF** Multimedia Resource Function; The unit responsible for media manipulation in IMS.

**MRFC** Multimedia Resource Function Controller; Controller part of the MRF of IMS.

**MRFP** Multimedia Resource Function Processor; Processor part of the MRF of IMS.

**MSIN** Mobile Subscriber Identification Number; Phone numbers to be called, belong to users.

**NB** Narrowband; A channel or medium that provides low speed data transport and/or process.

**NGN** Next Generation Network; A PS based network architecture that handles all conventional CS network services like PSTN voice, SMS etc.

**LTE** Long Term Evolution; The brand name for 4G mobile networks.

**P2P** Peer to Peer; Direct or tunneled end-to-end communication between communicating parties.

**PCEF** Policy and Charging Enforcement Function; The unit controls charging related matters in IMS.

**P-CSCF** Proxy Call Session Control Function; A SIP proxy, which is the first contact point for the IMS terminals (UEs), located both in the visiting and in the home network.

**PCM** Pulse Code Modulation; A method to digitize sampled analog signals.

**PDN** Packet Data Network; A connectionless network, in which the data is carried via small chunks.

**PGW** Packet Data Network Gateway; The gateway which transfers user plane data between RAN, IMS and other networks.

**PIN** Personal Identification Number; A secret code used for authentication and authorization.

**PS** Packet Switching; A communication methodology, implies transfer of data as small independent segments between peers.

**PSTN** Public Switched Telephone Network; The legacy CS based fixed line phone network.

**QoS** Quality of Service; A set of indicators about service quality.

**RAN** Radio Access Network; A part of mobile networks, connects UEs to core networks.

**RFID** Radio Frequency Identification; A usage of electromagnetic fields to wirelessly transfer data.

**RTP** Real Time Protocol; A protocol used to transport data as a real-time stream.

**RTCP** Real Time Control Protocol; A protocol used to deliver control messages for real-time streams.

**RTMP** Real Time Messaging Protocol; A media streaming protocol utilizes TCP.

**SGW** Signaling Gateway, The gateway which transfers control plane data between RAN, IMS and other networks.

**SIP** Session Initiation Protocol; A signaling protocol to initiate, control and terminate sessions of multimedia services.

**SMS** Short Messaging Service; A text messaging service in mobile networks.

**TISPAN** Telecommunications and Internet Converged Services and Protocols for Advanced Networking; A research body to move CS services to the PS domain.

**UE** User Equipment; User side mobile communication devices like phones, tablets etc.

**UICC** Universal IC Card, A chip equipped smart card that holds subscriber information and specific applications. Formerly known as the SIM card.

**UMTS** Universal Mobile Telecommunications System; The 3$^{rd}$ generation mobile cellular system for networks based on GSM.

**USIM** Universal Subscriber Identity Module; An improved SIM card application for LTE and UMTS.

**USSD** Unstructured Supplementary Services Data; A session based text oriented communication method between clients and MNOs over signaling channels in mobile networks.

**USSI** USSD Simulation Service in IMS; The way to support USSD within IMS framework.

**VoIP** Voice over Internet Protocol; A protocol built to carry digitized voice data over IP networks.

**WB** Wideband; A channel or medium that provides high speed data transport and/or process.

**XML** Extensible Markup Language; A portable markup language for text based information, designated to be both human and machine readable.

** All referenced internet links, including the ones given in the figure references and the appendices, are accessed and noted as valid, as of 20.02.2015.*

# Figure References

# Appendix A: Some of the LTE, UMTS and IMS Interfaces

| Interface Name | IMS Entities | Description | Protocol |
|---|---|---|---|
| Cr | MRFC, AS | Used by MRFC to fetch documents (e.g. scripts, announcement files, and other resources) from an AS. Also used for media control related commands. | TCP/SCTP channels |
| Cx | (I-CSCF, S-CSCF), HSS | Used to send subscriber data to the S-CSCF; including Filter criteria and their priority. Also used to furnish CDF and/or OCF addresses. | Diameter |
| Dh | AS (SIP AS, OSA, IM-SSF) <-> SLF | Used by AS to find the HSS holding the User Profile information in a multi-HSS environment. DH_SLF_QUERY indicates an IMPU and DX_SLF_RESP return the HSS name. | Diameter |
| Dx | (I-CSCF or S-CSCF) <-> SLF | Used by I-CSCF or S-CSCF to find a correct HSS in a multi-HSS environment. DX_SLF_QUERY indicates an IMPU and DX_SLF_RESP return the HSS name. | Diameter |
| Gm | UE, P-CSCF | Used to exchange messages between SIP user equipment (UE) or Voip Gateway and P-CSCF | SIP |
| Go | PDF, GGSN | Allows operators to control QoS in a user plane and exchange charging correlation information between IMS and GPRS network | COPS (Rel5), Diameter (Rel6+) |
| Gq | P-CSCF, PDF | Used to exchange policy decisions-related information between P-CSCF and PDF | Diameter |
| Gx | PCEF, PCRF | Used to exchange policy decisions-related information between PCEF and PCRF | Diameter |
| Gy | PCEF, OCS | Used for online flow based bearer charging. Functionally equivalent to Ro interface | Diameter |
| ISC | S-CSCF <-> AS | Reference point between S-CSCF and AS. Main functions are to :<br>• Notify the AS of the registered IMPU, registration state and UE capabilities<br>• Supply the AS with information to allow it to execute multiple services<br>• Convey charging function addresses | SIP |
| Ici | IBCFs | Used to exchange messages between an IBCF and another IBCF belonging to a different IMS network. | SIP |
| Izi | TrGWs | Used to forward media streams from a TrGW to another TrGW belonging to a different IMS network. | RTP |
| Ma | I-CSCF <-> AS | Main functions are to:<br>• Forward SIP requests which are destined to a Public Service Identity hosted by the AS<br>• Originate a session on behalf of a user or Public Service Identity, if the AS has no knowledge of a S-CSCF assigned to that user or Public Service Identity<br>• Convey charging function addresses | SIP |
| Mg | MGCF -> I,S-CSCF | ISUP signaling to SIP signaling and forwards SIP signaling to I-CSCF | SIP |
| Mi | S-CSCF -> BGCF | Used to exchange messages between S-CSCF and BGCF | SIP |
| Mj | BGCF -> MGCF | Used for the interworking with the PSTN/CS Domain, when the BGCF has determined that a breakout should occur in the same IMS network to send SIP message from BGCF to MGCF | SIP |
| Mk | BGCF -> BGCF | Used for the interworking with the PSTN/CS Domain, when the BGCF has determined that a breakout should occur in another IMS network to send SIP message from BGCF to the BGCF in the other network | SIP |
| Mm | I-CSCF, S-CSCF, external IP network | Used for exchanging messages between IMS and external IP networks | SIP |
| Mn | MGCF, IM-MGW | Allows control of user-plane resources | H.248 |
| Mp | MRFC, MRFP | Allows an MRFC to control media stream resources provided by an MRFP. | H.248 |
| Mr<br>Mr' | S-CSCF, MRFC<br>AS, MRFC | Used to exchange information between S-CSCF and MRFC<br>Used to exchange session controls between AS and MRFC<br>Application Server sends SIP message to MRFC to play tone and announcement. This SIP message contains sufficient information to play tone and announcement or provide information to MRFC, so that it can ask more information from Application Server through Cr Interface. | SIP |
| Mw | P-CSCF, I-CSCF, S-CSCF, AGCF | Used to exchange messages between CSCFs. AGCF appears as a P-CSCF to the other CSCFs | SIP |
| Mx | BGCF/CSCF, IBCF | Used for the interworking with another IMS network, when the BGCF has determined that a breakout should occur in the other IMS network to send SIP message from BGCF to the IBCF in the other network | SIP |
| P1 | AGCF, A-MGW | Used for call control services by AGCF to control H.248 A-MGW and Residential Gateways | H.248 |
| P2 | AGCF, CSCF | Reference point between AGCF and CSCF. | SIP |
| Rc | MRB, AS | Used by the AS to request that media resources be assigned to a call when utilizing MRB In-Line mode or In Query mode | SIP, In Query mode (Not specified) |
| Rf | P-CSCF, I-CSCF, S-CSCF, BGCF, MRFC, MGCF, AS | Used to exchange offline charging information with CDF | Diameter |
| Ro | AS, MRFC, S-CSCF | Used to exchange online charging information with OCF | Diameter |
| Rx | P-CSCF, PCRF | Used to exchange policy and charging related information between P-CSCF and PCRF<br>Replacement for the Gq reference point. | Diameter |
| S1-MME | eNodeB, MME | Reference point for the control plane protocol between E-UTRAN and MME. | S1AP, SCTP |
| S1-U | eNodeB, SGW | Reference point between E-UTRAN and Serving GW for the per bearer user plane tunnelling and inter eNodeB path switching during handover. | GTPU, UDP |
| SGi | PGW, IP Networks | It is the reference point between the PDN GW and the packet data network. Packet data network may be an operator external public or private packet data network or an intra-operator packet data network, e.g. for provision of IMS services. This reference point corresponds to Gi for 3GPP accesses. | Radius, Diameter |
| Sh | AS (SIP AS, OSA SCS), HSS | Used to exchange User Profile information (e.g., user related data, group lists, user service related information or user location information or charging function addresses (used when the AS has not received the third party REGISTER for a user)) between an AS (SIP AS or OSA SCS) and HSS. Also allow AS to activate/deactivate filter criteria stored in the HSS on a per subscriber basis | Diameter |
| Si | IM-SSF, HSS | Transports CAMEL subscription information including triggers for use by CAMEL based application services information. | MAP |
| Sr | MRFC, AS | Used by MRFC to fetch documents (scripts and other resources) from an AS | HTTP |
| Ut | UE and SIP AS (SIP AS, OSA SCS, IM-SSF) PES AS and AGCF | Facilitates the management of subscriber information related to services and settings | HTTP(S), XCAP |
| z | POTS, Analog phones and VoIP Gateways | Conversion of POTS services to SIP messages | |

# Appendix B: Call Flows for IMS-to-IMS and IMS-to-PSTN Calls

Call flow diagrams are generated by EventHelix.com Inc. (www.eventhelix.com). More call flows may be obtained from their website when necessary.

Standardized and recommended schemes are considered, real implementations of MNOs may differ from the figures.

For the best readability, use of digital version of this thesis is advised.